\documentclass[aoas]{imsart}
\RequirePackage{amsthm,amsmath,amsfonts,amssymb}
\RequirePackage[authoryear]{natbib}
\RequirePackage[colorlinks,citecolor=blue,urlcolor=blue]{hyperref}
\RequirePackage{graphicx}
\usepackage{multirow}
\usepackage{algorithm} 
\usepackage{algpseudocode}
\usepackage{color}
\usepackage[textfont = it]{caption}
\usepackage[textfont = it]{subcaption}
\startlocaldefs

\theoremstyle{remark}

\endlocaldefs
\begin{document}

\begin{frontmatter}
\title{A Flexible Zero-Inflated Poisson-Gamma Model with application to microbiome read counts}

\runtitle{Zero-Inflated Poisson-Gamma Model}
\begin{aug}

\author[A,B]{\fnms{Roulan} \snm{Jiang}},
\author[C,D]{\fnms{Xiang} \snm{Zhan}\ead[label=e2,mark]{zhanx@bjmu.edu.cn}}
\and
\author[A,B]{\fnms{Tianying} \snm{Wang}\ead[label=e3,mark]{tianyingw@tsinghua.edu.cn}}

\address[A]{Center for Statistical Science, Tsinghua University, Beijing 100084, China.
\printead{e3}}
\address[B]{Department of Industrial Engineering, Tsinghua University, Beijing 100084, China.}
\address[C]{Department of Biostatistics, School of Public Health, Peking University, Beijing 100191, China.
}
\address[D]{Beijing International Center for Mathematical Research, Peking University, Beijing 100871, China.
\printead{e2}}
\end{aug}

\begin{abstract}
\begin{center}
    Abstract
\end{center}
In microbiome studies, it is of interest to use a sample from a population of microbes, such as the gut microbiota community, to estimate the population proportion of these taxa. However, due to biases introduced in sampling and preprocessing steps, these observed taxa abundances may not reflect true taxa abundance patterns in the ecosystem. Repeated measures including longitudinal study designs may be potential solutions to mitigate the discrepancy between observed abundances and true underlying abundances. Yet, widely observed zero-inflation and over-dispersion issues can distort downstream statistical analyses aiming to associate taxa abundances with covariates of interest. 
To this end, we propose a Zero-Inflated Poisson Gamma (ZIPG) framework to address the aforementioned challenges. From a perspective of measurement errors, we accommodate the discrepancy between observations and truths by decomposing the mean parameter in Poisson regression into a true abundance level and a multiplicative measurement of sampling variability from the microbial ecosystem. Then, we provide flexible modeling by connecting both mean abundance and the variability to different covariates, and build valid statistical inference procedures for both parameter estimation and hypothesis testing. Through comprehensive simulation studies and real data applications, the proposed ZIPG method provides significant insights into distinguished differential variability and abundance.
\end{abstract}

\end{frontmatter}
\newpage
\section{Introduction} \label{sec:intro}

The human microbiome consists of the collection of all microbes living in or on the human body and plays an important role in maintaining human health \citep{manor2020health}. Quantification of the microbiome usually proceeds by 16s rRNA sequencing or metagenomic shotgun sequencing, where sequence read counts are often summarized into a taxa count table. Here the word \textit{taxa} generically refers to features such as operational taxonomic units or other taxonomic or functional groupings of bacterial sequences. A crucial task in microbiome research is to link these taxa counts to covariates of interest (e.g., disease status, health outcomes, and environmental conditions) via statistical analysis \citep{li2015microbiome}. To achieve this goal, one needs first to address some common challenges, such as zero inflation and over-dispersion in observed taxa counts, and more importantly, the discrepancy between observed taxa abundances in samples and true abundances in the underlying microbial ecosystem, such as the gut microbiota community, to guarantee rigor and reproducibility of the analysis results \citep{willis19}.

Owing to biases introduced in sampling extraction, polymerase chain reaction (PCR) amplification, sequencing, bioinformatics prepossessing, and other possible experimental procedures, observed taxa abundances may not well reflect unobserved true abundances in the ecosystem. While multiple versatile statistical methods have been proposed to address the aforementioned issue for microbiome compositional data \citep{shi2022high,martin2020modeling}, measurement errors actually occur at latent count variables rather than proportions. Such a compositional transformation may lose some variation/dispersion information that is important to subsequent statistical analysis \citep{mcmurdie2014waste,xu2021zero,li21}. Moreover, recent research indicates that it is possible to quantify microbial load (and hence the absolute abundance of each taxon) using flow cytometry \citep{vandeputte17}. Following this research vein, we will propose valid statistical inference for microbiome count data accommodating the discrepancy between observed sample abundances and underlying true abundances. Specifically, motivated by a recent inference procedure based on multiple rarefaction-based re-samplings \citep{hu21}, we take samples with repeated measures (or longitudinal measurements) to account for sampling fluctuations.

Like many high-throughput DNA sequencing assays exhibiting high sparsity, microbiome experiments often have about 50\% or more zero measurements
\citep{silverman20}. There are, in general, two types of approaches to handle these zeros in microbiome sequence count data. One is to impute zeros based on missing data scheme assumptions \citep{martin11} or random matrix low-rank assumptions \citep{cao20}. The imputation approach is often coupled with downstream log-ratio-based compositional data analysis.  The other approach is to propose a two-part model with a point probability mass at zero along with another parametric distribution. Examples include zero-inflated Poisson, zero-inflated negative binomial and many others \citep{xu2021zero,zhang2020fast,li18,tang19,zeng22}. While both approaches are popular in microbiome data analysis, a recent study demonstrates that a potential limitation of imputing zeros is that violation of underlying assumptions may distort downstream statistical analysis \citep{silverman20}. To this end, we will take the two-part modeling strategy to handle excessive zeros in microbiome data in this paper.

Suppose samples are collected from $n$ subjects, and every subject could have multiple measurements with counts of $K$ taxa measured in each sample. For $i=1,...,n$, $j=1,...,n_i$ and $k=1,...,K$, denote $W_{ijk}$ as the count of the $k$th taxon in the $j$th sample/measurement from the $i$th subject. For ease of presentation, we assume hereafter that sample $j$, with $j=1,...,n_i$,  was collected longitudinally from subject $i$, without loss of generality. The sequencing depth, or library size, of each sample is $M_{ij}=\sum_{k=1}^K W_{ijk}$. To account for the aforementioned excessive zeros issue, the zero-inflated Poisson distribution has been proposed to model microbiome counts \citep{xu2021zero}:
\begin{equation}\label{eq:poisson}
W_{ijk} \sim \left\{ \begin{matrix}
0 &{\rm with \ probability}\ p_{ijk}\\
{\rm Poisson} (\lambda_{ijk})  &{\rm \ \ with \ probability}\ 1-p_{ijk} ,
\end{matrix}\right.
\end{equation}
where $\lambda_{ijk}$ represents the mean abundance of taxon $k$ on the $j$th observation from subject $i$, and $p_{ijk}$ is the probability mass to model excessive zeros. A major critique of Poisson models is the failure to accommodate over-dispersion, which has been widely observed for sequence count data, including microbiome data. An alternative of Poisson is the negative binomial distribution, originally proposed for RNA sequence count data \citep{edgeR2010,DESeq2014} and recently extended to microbiome data \citep{zhang2020fast}. The zero-inflated negative binomial (ZINB) distribution is given by:
\begin{equation} \label{eq:ZINB}
W_{ijk} \sim \left\{ \begin{matrix}
0 &{\rm with \ probability}\ p_{ijk}\\
{\rm NB}(\mu_{ijk},\alpha_{k})  &{\rm \ \ with \ probability}\ 1-p_{ijk},
\end{matrix}\right.
\end{equation}
where $\lambda_{ijk}$ and $\alpha_k$ are the mean parameter and (over-)dispersion parameter of the negative binomial distribution, respectively. ZINB can also be expressed as a Gamma prior upon $\lambda_{ijk}$ in ZIP with $E(\lambda_{ijk}) = \mu_{ijk}$ and $Var(\lambda_{ijk}) = \mu_{ijk}^2 \alpha_{k}$. That is, 
$
    \lambda_{ijk} \sim {\rm Gamma}\left(\alpha_{k}^{-1},\mu_{ijk}\alpha_{k}\right),
$ where $\alpha_k$ is a nuisance parameter depending on the $k$th taxon only.

\begin{figure}[!ht]
\centering
\begin{subfigure}{0.49\textwidth}
  \centering
\includegraphics[width = 0.8\textwidth]{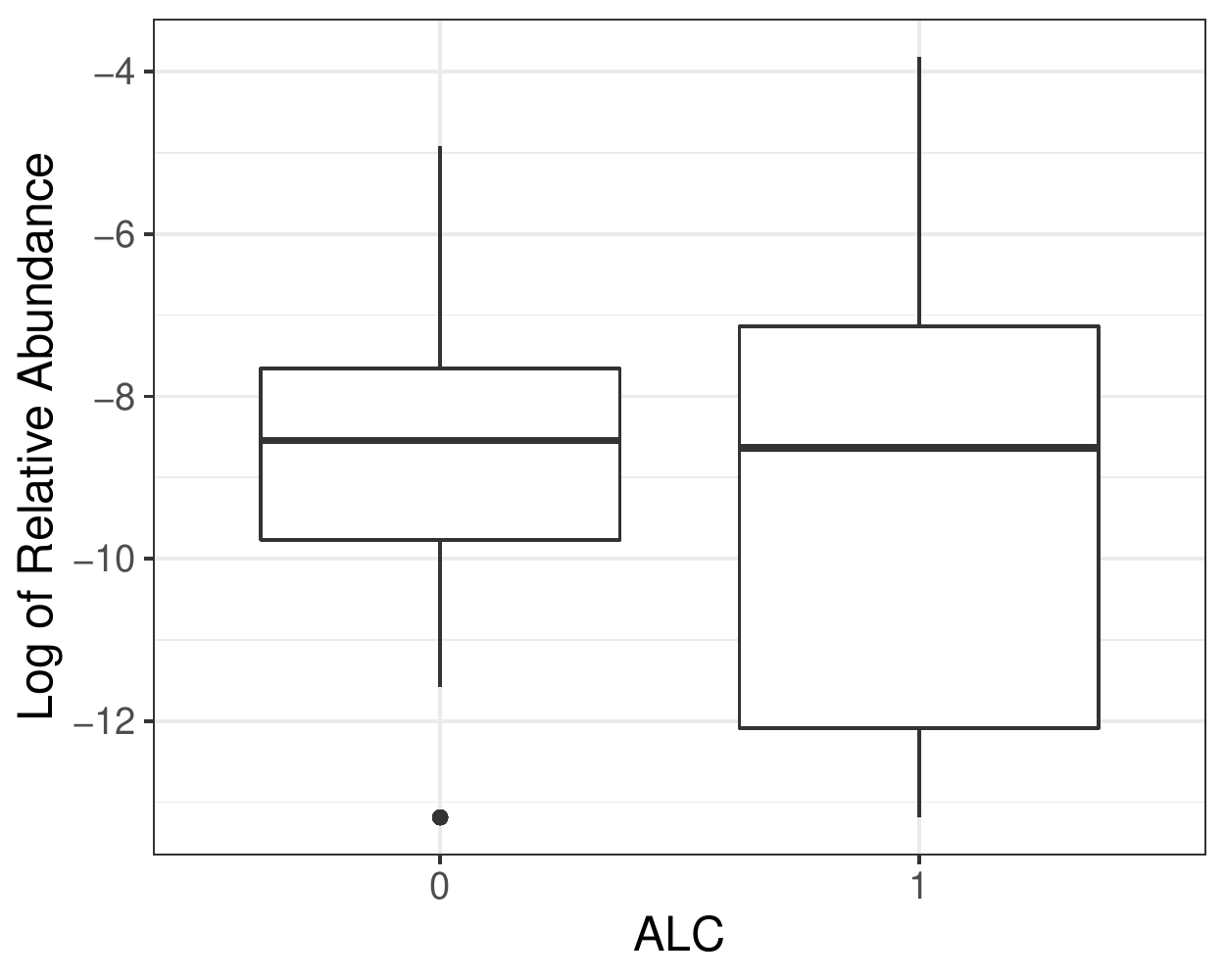}
  \caption{Raw data}
\end{subfigure}
\begin{subfigure}{0.49\textwidth}
  \centering
\includegraphics[width = 0.8\textwidth]{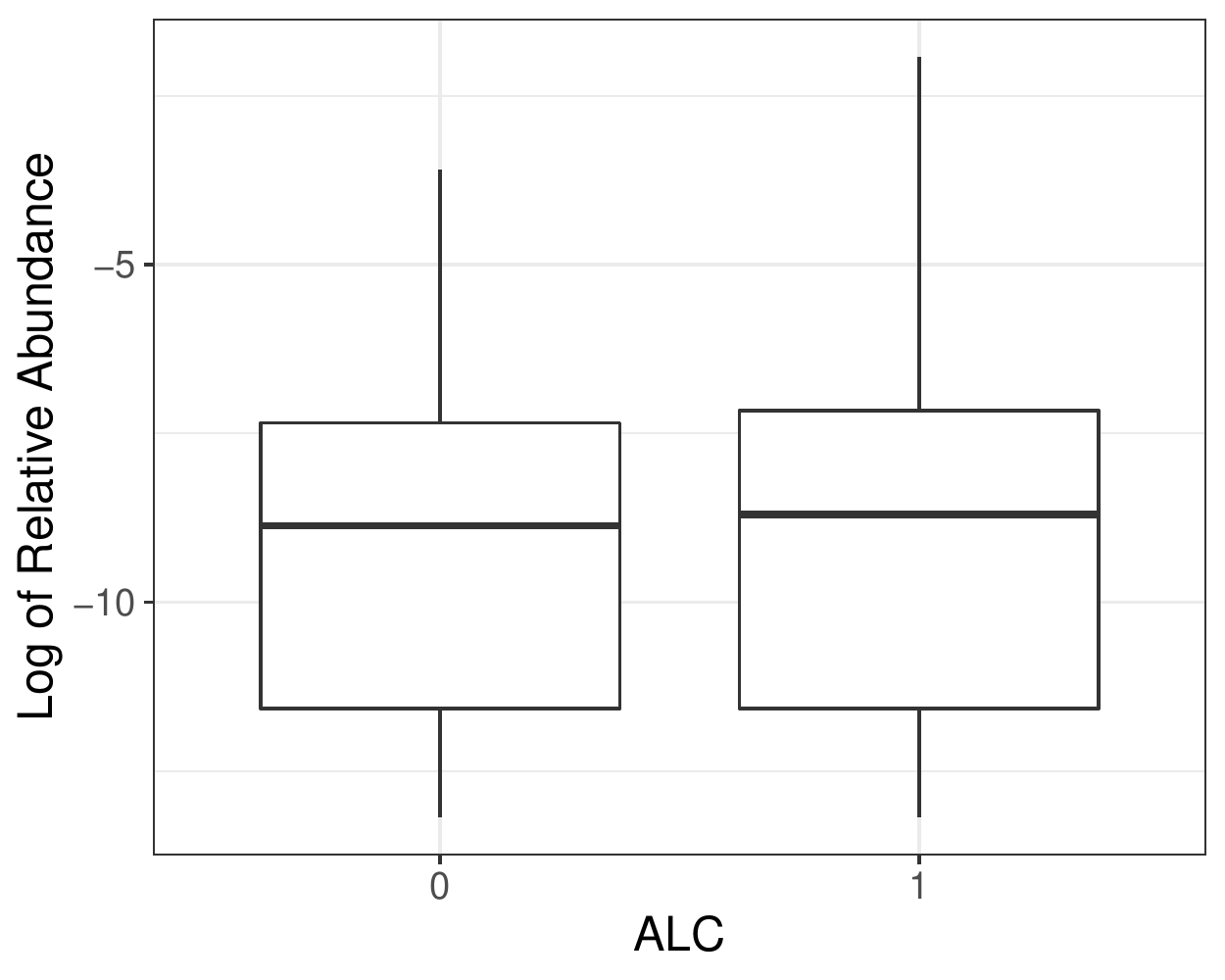}
  \caption{pscl predicted data}
\end{subfigure}
\caption{Box plot for log of relative abundance of Burkholderiales bacterium in alcohol (ALC = 1) and non-alcohol (ALC = 0) groups. }
\label{fig:Motivating_Example}
\end{figure}

Host factors, like disease status and dietary regimes, can impact microbiome stability, referred to as \textit{dysbiosis} to describe the imbalance of microbiome community during some unhealthy conditions compared to normal ones \citep{petersen2014defining}. Thus, it is of our interest to investigate the relationship between the taxa abundance variation and covariates, which is naturally caused by microbiome stability perturbation yet overlooked in the ZINB model. For illustration purposes, we use the taxon Burkholderiales bacterium from the diet-microbiome study \citep{johnson2019daily} to show potential abundance variation associated with covariates. One primary goal of the diet-microbiome study is to analyze the microbial difference between alcohol drinkers and teetotalers. According to box plots for the raw data and the predicted distribution using the ZINB model by \texttt{R} package pscl \citep{Rpscl_zeroinfl}, we observe that two groups have similar mean abundances, but the variance among alcohol drinkers is evidently larger, which can not be captured by pscl (Figure \ref{fig:Motivating_Example}). Therefore, it is crucial to consider variation in microbiome count data to obtain accurate and robust analysis results.

To address potential limitations, we propose a Zero-Inflated Poisson-Gamma (ZIPG) framework, which provides flexible modeling by connecting both the mean abundance and its dispersion with different sets of covariates, respectively. First, we consider a hierarchical model by adjusting the mean parameter $\lambda_{ijk}$ of ZIP (i.e., model \eqref{eq:poisson}) with a multiplicative factor $U_{ijk}$, whose distribution is Gamma and can be viewed as a multiplicative measurement error. Further, we construct the ZIPG model and connect mean parameter $\lambda_{ijk}$ and variation factor $U_{ijk}$ to different sets of covariates. Note that our model is different from the traditional Bayesian expression of negative-binomial in the sense that we model the mean and variability of taxa abundance separately, providing a meaningful explanation from the individual-level variation perspective.

Our contributions are two-fold. First, our ZIPG framework provides flexible modeling of microbiome sequence counts with repeated measures and allows us to analyze how different sets of variables affect both  mean taxa abundance and its dispersion. Second, within the ZIPG framework, we develop inference procedures, including point and interval estimates and hypothesis testing, to examine the relationship between microbial taxa abundances and covariates of interest. By introducing the variation factor as a multiplicative measurement error term, our ZIPG method is able to capture higher-order moment information of taxa abundance and has been shown to be more powerful than ZINB-based methods. Through extensive simulations, we illustrate that existing ZINB-based methods could have severely inflated type I error when differential variability exists, whereas ZIPG can control type I error around the nominal level. When analyzing two real microbiome data sets, ZIPG identified more significant taxa than two ZINB models under the same nominal false discovery rate level, and also distinguished differential variability and differential abundance, providing more insights for further biological or biomedical functional investigations.

The rest of this paper is organized as follows. In Section~\ref{s:model}, we introduce our ZIPG model and discuss some parameters of interest. ZIPG model fitting and hypothesis testing procedures are proposed in Section~\ref{s:method}. In Section~\ref{s:simulation}, we demonstrate the superior performance of our approach under different simulation settings. We apply our method to the data from a vaginal microbiome study and a diet-microbiome study in Section~\ref{s:application}, and conclude it with a brief discussion in Section~\ref{s:discussion}.

\section{Model and Notation}\label{s:model}
\subsection{Zero-Inflated Poisson Gamma Model}
For taxon count $W_{ijk}$, we decompose the mean of Poisson distribution in eq \eqref{eq:poisson} into a true abundance level $\lambda_{ijk}$ and a multiplicative factor $U_{ijk}$ and consider the following hierarchical Zero-Inflated Poisson-Gamma (ZIPG) model:
\begin{eqnarray}\label{eq:ZIPG}
 {W}_{ijk} \mid U_{ijk} &\sim& \left\{ \begin{matrix}
 \phantom{-}0 &\phantom{-}{\rm with \ probability}\ p_k\\
 {\rm Poisson}(\lambda_{ijk}U_{ijk})\  &{\rm with \ probability}\ 1-p_k,\\
 \end{matrix}\right. \\
  U_{ijk} &\sim& {\rm Gamma}(\theta_{ik}^{-1},\theta_{ik}),\nonumber
\end{eqnarray}
where $\lambda_{ijk}$ represents the true abundance level for taxon $k$ on the $j$th observation from subject $i$, and $p_{k}$ denotes the zero-inflation parameter describing the probability of true zero occurrence of taxon $k$. $U_{ijk}$ follows Gamma distribution with the same rate parameter and shape parameter $\theta_{ik}^{-1}$.  This factor does not change the average abundance level given the fact that $E(U_{ijk})=1$. On the other hand, $Var(U_{ijk}) = \theta_{ik}$ allows extra variation of the observed abundance $W_{ijk}$ around the average level $\lambda_{ijk}$, a phenomenon described as the deviation of observed abundance to unobserved true abundance. It is of note that the variability term, $\theta_{ik}$, remains the same for all measurements (i.e., $j=1,...,n_i$) across individual $i$. Thus it reflects the stability of taxon $k$ in the microbial system of individual $i$. This is motivated by the assumption about metagenomic sequencing bias being taxon-specific but not sample-specific made in literature \citep{mclaren2019consistent}. Finally, we assume the zero-inflation parameter $p_k$ is only taxon-specific and is common across samples ($j$) and individuals ($i$). This is because many experimental factors in sequencing can introduce the measurement of zeros \citep{silverman20} and hence it is less appealing to link zero inflation parameter $p_k$ to other covariates (e.g., biological or environmental conditions) possessed by individuals or samples. We have checked the sensitivity of ZIPG under model misspecification when this assumption is violated by comparing the performance of the current ZIPG model \eqref{eq:ZIPG} to a full ZIPG model (denoted as ZIPG-full), which replaces $p_k$ by $p_{ijk}$ and links $p_{ijk}$ to covariates. Results in Section \ref{s:Model-sensitivity} indicates that the current ZIPG model tends to have better model-fitting performance than ZIPG-full, which is consistent with previous empirical conclusions that modeling zero inflation in sequence count data should be careful (with respect to underlying zero generating process) and numerical evidence tends to favor simpler models \citep{silverman20}.

To further explore the new ZIPG framework, {\color{black}let $W_{ijk}^{\rm Pois} \sim {\rm Poisson}(\lambda_{ijk}) $ denote as the random variable generated from the Poisson distribution and $W_{ijk}^{\rm PG}$ as the random variable generated from the Poisson-Gamma part in eq~\eqref{eq:ZIPG}. We have
\begin{eqnarray*}
   && E(W_{ijk}^{\rm PG}) = E(W_{ijk}^{\rm Pois}) = \lambda_{ijk}, \\
    &&Var(W_{ijk}^{\rm PG}) =   \lambda_{ijk}(1+\lambda_{ijk}\theta_{ik}) =Var(W_{ijk}^{\rm Pois})(1+\lambda_{ijk}\theta_{ik}).
\end{eqnarray*}}
Thus, the mean of the Poisson-Gamma distribution is the same as the regular Poisson distribution, but its variance is multiplied by $(1+\lambda_{ijk}\theta_{ik})$ to account for the over-dispersion caused by the multiplicative measurement error factor $U_{ijk}$.  We also observe the similar phenomenon of using a more sophisticated hierarchical model to account for over-dispersion in microbiome data analysis, such as the Beta-Binomial distribution \citep{martin2020modeling} and Dirichlet-multinomial distribution \citep{la12}. In this paper, we refer to $\lambda_{ijk}$ as the abundance mean parameter and $\theta_{ik}$ as the abundance dispersion parameter.

\subsection{Parameters of Interest}
A critical task in microbiome research is to explore the relationship between taxa abundances and covariates of interest. Compared to noisy observed counts $W_{ijk}$, it is more interesting to investigate the association between key parameters (i.e., $\lambda_{ijk}$ and $\theta_{ik}$) of the underlying taxa abundance distribution and covariates of interest. To this end, we connect the mean parameter and dispersion parameter with different sets of covariates, respectively. In the repeated measures or longitudinal study design considered in the current paper, some covariates vary across different samples within the same subject,  such as dietary intake, and we refer to them as ``time-dependent" covariates. Other covariates, which do not change during the study, such as the treatment group assigned at the beginning of the study, are referred to as ``time-independent" covariates. {\color{black}Since the dispersion parameter $\theta_{ik}$ describes the deviation of  short-term abundance from long-term mean abundance $\lambda_{ijk}$, we propose to link it to time-independent covariates, supported by the evidence of microbiome stability perturbation in \cite{morgan2012dysfunction} and \cite{couch2021diet}.} For the mean abundance $\lambda_{ijk}$, it can be linked to either time-dependent covariates or time-independent covariates, or both. Therefore, we define the following link functions:
\begin{equation}\label{eq:ZIPG_lambda}
g(\lambda_{ijk})  =  \beta_{k,0} + \boldsymbol{X_{ij}}^T\boldsymbol{\beta_k} + \log (M_{ij}), \quad 
 g^*(\theta_{ik})  = \beta^{*}_{k,0} + \boldsymbol{X^{*T}_{i}} \boldsymbol{\beta^{*}_k},
\end{equation}
where $\boldsymbol{X_{ij}} \in \mathbb{R}^{d_1}$ is a vector of covariates associated with $\lambda_{ijk}$, $\boldsymbol{X^*_{i}}\in \mathbb{R}^{d_2}$ include covariates associated with $\theta_{ik}$, $\boldsymbol{\beta_k} = (\beta_1,...,\beta_{d_1})^T$ and $\boldsymbol{\beta_k^*} = (\beta_1^*,...,\beta_{d_2}^*)^T$ are regression coefficients of interest. We allow overlapped covariates in two models. For ease of presentation, we term the two models in \eqref{eq:ZIPG_lambda} as ZIPG mean model and ZIPG dispersion model, respectively. The  $\log(M_{ij})$ term in the mean model accounts for the effect of sequencing depth variation on mean abundances. The same offset is used in previous ZINB models \citep{zhang2020fast,edgeR2010}, and the log of median-of-ratios is another possible candidate for offset used in literature \citep{xu2021zero,DESeq2014}. While most existing ZINB-based methods \eqref{eq:ZINB} models $\mu_{ijk}$ but treat $\alpha_k$ as a nuisance parameter, our approach allows additional time-independent covariates linked to the dispersion of abundance, which would lead to better model fit and more powerful association analysis as will be shown later in this paper. According to previous discussions, we suggest including time-independent covariates such as demographic and lifestyle-related variables in $\boldsymbol{X^*_{i}}$. All covariates of interest, regardless of time-dependent or not, shall be included in $\boldsymbol{X_{ij}}$. By testing the coefficients $\boldsymbol{\beta_k}$ and $\boldsymbol{\beta^*_k}$, we can detect differential abundance and differential variability impacted by physiological status or host environment, respectively. Finally, if we do not include any covariates in $\boldsymbol{X^*_{i}}$, our ZIPG model will degenerate to ZINB  (i.e., model \eqref{eq:ZINB})  with $\theta_{ik} = \alpha_k$ for any subject.  Throughout this paper we choose a logarithmic link function $g(x) = g^*(x) := \log(x)$ to ensure $\lambda_{ijk} > 0$ and  $\theta_{ik}>0$.

The proposed hierarchical ZIPG model has several key advantages. First, to handle over-dispersion in microbiome count data, we decompose abundances into the long-term true abundance and its individual-level variation through a multiplicative factor. Second, we allow different sets of variables associated with the mean and the variation of abundance and provide explanations for variations in the individual-level microbial system. Thus, the proposed method is not only able to test the change of the mean abundance but also the microbiome stability affected by  physiological status or host environment, which cannot be achieved by existing ZIP models. In addition, parameter $\theta_{ik}$ also controls the skewness and kurtosis of the Gamma distribution. That is, we allow the higher-order moments (or the shape of the distribution) to be linked to covariates, which is another feature that is typically missed in existing models.

\section{Statistical Inference in ZIPG} \label{s:method}
In this section, we develop statistical inference procedures in ZIPG, including parameter point estimation, interval estimation, and hypothesis testing. For ease of presenting, we omit the subscript $k$ and simply denote the parameter set associated with taxon $k$ as $\boldsymbol{\Omega} = (\beta_0, \boldsymbol{\beta}^T, \beta^*_0, \boldsymbol{\beta^*}^T,\gamma)^T$, where $\gamma = \log\left\{p/(1-p)\right\}$ is the logit transformation of zero-inflated parameter $p$ in ZIPG model \eqref{eq:ZIPG}.

\subsection{Model Fitting} \label{s:Modelfitting}
Given covariates $\boldsymbol{X}$ and $\boldsymbol{X^*}$, observed count data $\boldsymbol{W}$ and sequencing depth $\boldsymbol{M}$, we write the log-likelihood of $\boldsymbol{\Omega}$ as follows:
\begin{eqnarray} \label{eq:W_loglik}
 L (\boldsymbol{\Omega}\mid \boldsymbol{W})
&=& \sum_{i=1}^{n}\sum_{j=1}^{n_i} \left[
I(W_{ij}=0)
\log \left\{ \exp(\gamma)+ P_{\rm PG}(W_{ij} \mid \boldsymbol{\Omega}) \right\}
\right. 
\nonumber\\
 &  &\hskip 2mm \left.+ I(W_{ij}>0)\log \left\{P_{\rm PG}(W_{ij} \mid \boldsymbol{\Omega})\right\} -\log
\left\{\exp(\gamma)+1\right\}
\right] ,\\
{\rm with\ \ }P_{\rm PG}(W_{ij} \mid\boldsymbol{\Omega}) &=&  
\frac{\Gamma(W_{ij}+\theta_{i}^{-1})}{\Gamma(W_{ij}+1) \Gamma(\theta_{i}^{-1})} \frac{(\lambda_{ij}\theta_{i})^{W_{ij}}}
{(1+\lambda_{ij}\theta_{i})^{\theta_{i}^{-1}+W_{ij}}}, 
 \nonumber 
\end{eqnarray}
where $\lambda_{ij}$ and $\theta_{i}$ are functions of $\boldsymbol{\Omega}$ defined in eq \eqref{eq:ZIPG_lambda} and $\Gamma(x) = \int_0^{+\infty} t^{x-1}\exp(-t) dt$ is the Gamma function. The log-likelihood eq \eqref{eq:W_loglik} is non-concave in $\boldsymbol{\Omega}$ (see Section 1.1 of Supplements). In practice, we found that directly maximizing eq \eqref{eq:W_loglik} can cause trouble in distinguishing zeros from the Poisson-Gamma part and the other zero-inflation part of the ZIPG model, leading to an unreasonably low estimator of $\gamma$. A similar phenomenon has been observed for pscl, with more discussions provided in Section \ref{s:simulation}. Therefore, we use the EM algorithm for a reliable estimator of $\boldsymbol{\Omega}$.

Let $z_{ij}$ be the latent variable, where $z_{ij}=1$ indicates $W_{ij}$ is generated from zero-inflated part with probability $p=\exp(\gamma)/(\exp(\gamma)+1)$, and $z_{ij}=0$ indicates $W_{ij}$ is generated from the Poisson-Gamma distribution with probability $1-p$. The log-likelihood with complete data $\{W_{ij},M_{ij},X_{ij},X_{ij}^*,z_{ij}\}$ for $i=1,...,n$ and $j=1,...,n_i$ is written as:
\begin{equation} \label{eq:W_loglik_EM}
L (\boldsymbol{\Omega}\mid \boldsymbol{W},\boldsymbol{z}) 
= \sum_{i,j}\left[
z_{ij} \log(p)  + (1-z_{ij})\log\left\{(1-p)P_{\rm PG}(W_{ij} \mid \boldsymbol{\Omega})\right\}
\right] .
\end{equation}
The detailed procedure of the EM algorithm is provided in Algorithm~\ref{algorithm:EM}.  We first initialize $\boldsymbol{\Omega}^{(0)}$ by the results of \texttt{zeroinfl} in \texttt{pscl} with $\boldsymbol{\beta^*}=\boldsymbol{0}$. $p_{ij}^{(0)}$ is adjusted to the proportion of observed zeros of $\boldsymbol{W}$ to avoid the local maximum at the start point. Then we can repeat the E-step and M-step until convergence or the maximum number of iterations $t_{\rm max}$ is reached. For the $t$-th iteration, in M-step, we update $\boldsymbol{\Omega}$ by maximizing eq \eqref{eq:W_loglik_EM} using BFGS in \texttt{R} function \texttt{optim}  (\cite{broyden1970convergence,fletcher1970new,goldfarb1970family,shanno1970conditioning}) given latent variable $\boldsymbol{z}^{(t-1)}$. The gradient of eq \eqref{eq:W_loglik_EM} is applied to improve the computational efficiency (see Section 1.2 of Supplements). 
\begin{eqnarray}\label{M-step}
   {\boldsymbol{\Omega}^{(t)}}=\mathop{\arg\max}_{\boldsymbol{\Omega}} L (\boldsymbol{\Omega}\mid \boldsymbol{W},\boldsymbol{z}^{(t-1)}).
\end{eqnarray}
In E-step, we update latent variables $z_{ij}$ by their conditional expectations given $\boldsymbol{\Omega}^{(t)}$ estimated from M-step:
\begin{eqnarray}\label{E-step}
&& {z}_{ij}^{(t)} = E\left\{{z}_{ij}\mid W_{ij},\boldsymbol{\Omega}^{(t)}\right\}\\
& = &
\frac{P(W_{ij}\mid{z}_{ij}=1,\boldsymbol{\Omega}^{(t)})P({z}_{ij}=1 \mid \boldsymbol{\Omega}^{(t)})}
{P(W_{ij}\mid{z}_{ij}=1,\boldsymbol{\Omega}^{(t)})P({z}_{ij}=1 \mid \boldsymbol{\Omega}^{(t)})+
P(W_{ij}\mid{z}_{ij}=0,\boldsymbol{\Omega}^{(t)})P({z}_{ij}=0 \mid \boldsymbol{\Omega}^{(t)})} \nonumber\\
& = & \frac{I(W_{ij}=0)p^{(t)}}{I(W_{ij}=0)p^{(t)}+P_{\rm PG}(W_{ij}\mid \boldsymbol{\Omega}^{(t)})(1-p^{(t)})} .\nonumber
\end{eqnarray}

Substituting latent variables $z_{ij}^{(t)} = z(W_{ij},\boldsymbol{\Omega}^{(t)})$ into eq \eqref{eq:W_loglik_EM}, we can write the expectation of log-likelihood on a single observation as a new function 
 \begin{eqnarray*}
  Q_{ij}(\boldsymbol{\Omega}\mid \boldsymbol{\Omega}^{(t)})
  &=& E\left\{ L(\boldsymbol{\Omega} \mid W_{ij}, z_{ij}) \mid W_{ij}, \boldsymbol{\Omega}^{(t)} \right\} \\
  &=& z(\boldsymbol{\Omega}^{(t)},W_{ij})\log p + \left\{1-z(\boldsymbol{\Omega}^{(t)},W_{ij})\right\}\log\left\{(1-p) P_{\rm PG}(W_{ij} \mid  \boldsymbol{\Omega})\right\},
\end{eqnarray*}
 then $Q(\boldsymbol{\Omega} \mid \boldsymbol{\Omega}^{(t)}) = \sum_{ij} Q_{ij}(\boldsymbol{\Omega}\mid \boldsymbol{\Omega}^{(t)})$ is the quantity maximized in M-step equivalently. 
 
According to Theorem~6 in \cite{wu1983convergence}, since $\partial Q(\boldsymbol{\Omega} \mid \boldsymbol{\Omega}^{*}) /\partial \boldsymbol{\Omega}$ is continuous in $\boldsymbol{\Omega}$ and $\boldsymbol{\Omega}^{*}$, and our algorithm ensures $\partial Q(\boldsymbol{\Omega} \mid \boldsymbol{\Omega}^{(t)}) / \partial \boldsymbol{\Omega} \mid_{\boldsymbol{\Omega} = \boldsymbol{\Omega}^{(t+1)}}=\textbf{0}$ in each iteration, then EM estimator will converge to a stationary point of $L (\boldsymbol{\Omega}\mid \boldsymbol{W})$. {\color{black}Under mild conditions, EM Algorithm converges to the local maximum \citep[Theorem~3]{wu1983convergence}. Based on our suggested initialization, namely the MLE assuming $\boldsymbol{\beta^*} = 0$ with adjusted $p^{(0)}$, numerical studies suggest that our estimators are nearly unbiased.
  }
 \begin{algorithm} [!ht]
	\caption{ZIPG Expectation Maximization Algorithm}  \label{algorithm:EM}
	\begin{algorithmic}
	\Require $\boldsymbol{W},\boldsymbol{M},\boldsymbol{X},\boldsymbol{X^*}$, the maximum iterations $t_{\max}$, a tolerance $\epsilon_{\rm tol}$.
	\State Initialize $\boldsymbol{\Omega}^{(0)}$ by adjusted pscl estimation regardless of $\boldsymbol{X^*}$, set $t=0$.
	\State Initialize $\boldsymbol{z}^{(0)}$ with $\boldsymbol{\Omega}^{(0)}$ based on eq~\eqref{E-step}.
	\State Calculate $L^{(0)} = L(\boldsymbol{\Omega}^{(0)}\mid \boldsymbol{W}, \boldsymbol{z}^{(0)})$ as eq \eqref{eq:W_loglik_EM}.
	\While{$t < t_{\max} \  \text{and} \  |L^{(t)}-L^{(0)}|/|L^{(0)}| <\epsilon_{\rm tol}$}
			\State Given  $\boldsymbol{z}^{(t-1)}$, estimate $\boldsymbol{\Omega}^{(t)}$ by eq \eqref{M-step} (M-step).
			\State Get the maximized $L^{(t)} = L(\boldsymbol{\Omega}^{(t)}\mid \boldsymbol{W}, \boldsymbol{z}^{(t-1)})$ as in eq \eqref{eq:W_loglik_EM}.
			\State  Given $\boldsymbol{\Omega}^{(t)}$, update $\boldsymbol{z}^{(t)}$ by eq \eqref{E-step} (E-step).
			\State $t = t+1$.
			\EndWhile
	\State \textbf{Return} $\boldsymbol{\Omega}^{(t)}$.
	\end{algorithmic} 
\end{algorithm}

\subsection{Hypothesis Testing}\label{s:hypothesis}

{\color{black} In this paper, proving the asymptotic normality of the EM estimator is less of our interest, and despite its lack of rigorous theoretical justification, the EM estimator has been essentially treated as MLE in literature. Thus, we treat EM estimator $\boldsymbol{\hat\Omega}$ as MLE and construct the Wald test statistics and confidence interval based on the asymptotic normality of MLE. Further, we carefully consider the potential practical issues and evaluate different bootstrap methods and interval construction strategies with extensive numerical studies.}

Consider the null hypothesis $H_0 : \boldsymbol{A\boldsymbol{\Omega} = b}$ as the general form to test arbitrary subsets and linear combinations of parameters within $\boldsymbol{\Omega}$, where $\boldsymbol{A} \in \mathbb{R}^{r\times(d_1+d_2+3)}$ has full rank, $r < d_1+d_2+3$, and $\boldsymbol{b} \in \mathbb{R}^{r}$. The classical Wald test statistic can be constructed as  
$$
    {T}_{\rm Wald} = N (\boldsymbol{A{\boldsymbol{\hat\Omega}}-b})^T 
    (\boldsymbol{A}
    \boldsymbol{V}
    \boldsymbol{A}^T)^{-1}
    (\boldsymbol{A{\boldsymbol{\hat\Omega}}-b}),
$$
{\color{black} where $\boldsymbol{V}$ is the asymptotic covariance matrix of $\boldsymbol{\Omega}$, $N$ is the sample size, $T_{\rm Wald}$ is asymptotic $\chi^2_r$ under the null hypothesis. In practice, we found that directly using the inverse of observed information derived from the last M-step might underestimate the variance matrix (see Section 3.1 of the Supplements).} Thus, we propose to use nonparametric bootstrap to estimate the covariance matrix and construct the test statistic as in Algorithm~\ref{algorithm:bWald}.

\begin{algorithm}
\caption{ZIPG Bootstrap Wald Test} \label{algorithm:bWald}
\begin{algorithmic}
\Require $\boldsymbol{W},\boldsymbol{M},\boldsymbol{X},\boldsymbol{X^*}$, bootstrap replicates $B$.
\State Estimate $\boldsymbol{\hat\Omega^0}$ by EM (Algorithm~\ref{algorithm:EM}).
	\For {$b = 1,\ldots,B$}
	\State Randomly draw samples {\color{black}regarding all measurements} from original data at the same sample size with replacement, thus we get $\boldsymbol{W^b},\boldsymbol{M^b},\boldsymbol{X^b},\boldsymbol{{X^*}^b}$. 		
	\State Estimate $\boldsymbol{\hat\Omega^b}$ using $\boldsymbol{W^b},\boldsymbol{M^b},\boldsymbol{X^b},\boldsymbol{{X^*}^b}$ by EM (Algorithm~\ref{algorithm:EM}).
		\EndFor
	\State Compute the covariance matrix of $\boldsymbol{\hat\Omega^b}$ as $\boldsymbol{\hat V} = Var(\boldsymbol{\hat\Omega^b})$ .
	\State Compute 
	$$    \hat{T}_{\rm Wald} = (\boldsymbol{A{\boldsymbol{\hat\Omega^0}}-b})^T 
    (\boldsymbol{A} \boldsymbol{\hat V} \boldsymbol{A}^T)^{-1}
    (\boldsymbol{A{\boldsymbol{\hat\Omega^0}}-b}).$$ 
	\end{algorithmic} 
\end{algorithm}

The 100$(1-\alpha)\%$ confidence interval for any single parameter $\beta \in \boldsymbol{\Omega}$ can also be obtained through nonparametric bootstrap Wald test as $\left(\hat\beta- z_{\alpha/2} {\rm SD}(\beta^{b}),\hat\beta + z_{\alpha/2} {\rm SD}(\beta^{b})\right)$ based on the standard error (SD) of  $\beta^{b} \in \boldsymbol{\Omega^{b}}$ from $B$ bootstrap samples.  When the sample size is small or when the collected data is unbalanced (e.g., Romero data in Section \ref{s:application}), parametric bootstrap could be applied for more robust results as in \cite{martin2020modeling}. Thus, we also developed the Wald test based on parametric bootstrap (i.e., ZIPG-pbWald). {\color{black}That is, we simulate bootstrap samples from the ZIPG model with its parameters estimated under $H_0$. A detailed algorithm is provided in Section 2 of the Supplements.} 

{\color{black}
There are multiple ways to construct test statistics and confidence intervals. We conducted extensive simulations studies evaluating the performance of (1) the Wald test without bootstrap (ZIPG-Wald) and the likelihood ratio test (ZIPG-LRT); (2) nonparametric and parametric bootstrap through different construction strategies, such as normality-based/quantile-based/$\rm BC_a$\citep{efron1994introduction} intervals; (3) resampling schemes for ZIPG-bWald, such as resampling based on measurements or subjects. Simulation results suggest that bootstrap-based Wald tests (i.e., ZIPG-bWald and ZIPG-pbWald) with normality-based confidence intervals are desired, and resampling based on measurements yields satisfactory results. More discussions are provided in Section \ref{sec:sim_additional}.}



\section{Simulation Studies} \label{s:simulation}
We have conducted comprehensive numerical studies to
evaluate the performance of ZIPG in terms of both hypothesis testing and point/interval estimation. We evaluated both ZIPG with bootstrap Wald test (denoted as ZIPG-bWald) and parametric bootstrap Wald test (denoted as ZIPG-pbWald),  and then we compared them with ZINB-based methods, including NBZIMM \citep{zhang2020fast} and pscl \citep{Rpscl_zeroinfl}. We also assessed the  Poisson-Gamma (PG) model implemented by \texttt{glmmtmb} \citep{brooks2017glmmtmb} with the Wald test (no bootstrap).  PG can link covariates to $\lambda$ and $\theta$ as in ZIPG, but it is not adjusted for inflated zeros. We summarized all methods compared in Section~\ref{s:simulation} in Table~\ref{tab:method}.

Besides hypothesis testing and estimation, we further check the sensitivity of proposed ZIPG inference procedures under misspecified models. In particular, we evaluated the performance of ZIPG when the true zero proportion $p$ is also associated with covariates (Section~\ref{s:Model-sensitivity}), and when data is generated from Poisson-Gamma distribution or zero-inflated Beta-Binomial distribution (see Section 3.5 of the Supplements). 

\begin{table}[!ht]
    \centering
\caption{Summary of methods evaluated in simulation studies.}
    \begin{tabular}{c c c c  c}
    \hline
Method & $\lambda$ & $\theta$ & Zero-Inflation  & Notes\\
\hline
         ZIPG (proposed) & $\checkmark$ & $\checkmark$ & $\checkmark$  &  (parametric) bootstrap-based test\\
         pscl & $\checkmark$ & $\times$ & $\checkmark$ \\
         NBZIMM & $\checkmark$ & $\times$ & $\checkmark$  & consider random effect on $\lambda$\\
         PG & $\checkmark$ & $\checkmark$ & $\times$ \\
         
    \hline
    \end{tabular}
    \label{tab:method}
\end{table}

\subsection{Simulation Settings} 
We simulated $n=20$ subjects with $m =\{5,25\}$ measurements for each subject, with a total sample size of $N =\{100,500\}$. For $i =1,...,n$ and $j = 1,...,m$, we generated covariates $\boldsymbol{X_{ij}} = (X_{i,1}, X_{ij,2})$ associated with the mean parameter $\lambda_{ij}$ and $\boldsymbol{X^*_i} = X_{i,1}$ associated with the dispersion parameter $\theta_i$, where $X_{i,1}$ was a time-independent indicator sampled from a Bernoulli distribution with equal probabilities and served as the group indicator for each subject, such as pregnant or non-pregnant, alcohol drinker or non-alcohol drinker. $X_{ij,2}$ was a time-dependent longitudinal measurement generated from $X_{ij,2} = X_{i,2} + \epsilon_{ij}$, where $X_{i,2} \sim \mathcal{N}(0,1)$ and $\epsilon_{ij} \sim \mathcal{N}(0,0.1)$, representing covariates with variation in different measurements, such as calorie intake. Sequencing depths $M_{ij}$'s were generated based on the empirical distribution of observed sequencing depths in the Romero data set \citep{romero2014composition} analyzed later in Section~\ref{s:application}. Finally, the observed count data $\boldsymbol{W}$ was generated based on models \eqref{eq:ZIPG}-\eqref{eq:ZIPG_lambda} with $\boldsymbol{\beta}$ and $\boldsymbol{\beta^*}$ specified as below. 

To imitate real-world microbial data, we set  $(\beta_0, \beta^*_0) = (-4.23,0.6)$, guided by the ZIPG estimates in Romero data. We investigate performance of ZIPG under different model parameter configurations (Table \ref{table:simulation_setting}). In each simulation setting, we explored three different values of the zero-inflation proportion $p = \{ 0.3, 0.5, 0.7\}$, which is equivalent to $\gamma = \{-0.847, 0, 0.847\}$. For power analysis, we set $p = 0.5$ and vary the parameter of interest under each null hypothesis from 0 to 1.8. We compare ZIPG-bWald, ZIPG-pbWald, pscl, NBZIMM, and PG for the inference of $\beta_1$ and only ZIPG-bWald, ZIPG-pbWald, and PG for the inference on $\beta^*_1$, because inference of the dispersion parameter is not applicable in pscl and NBZIMM. Results are presented based on $L=1,000$ Monte Carlo replicates for each scenario. The performances of ZIPG-bWald and ZIPG-pbWald are evaluated using $B=200$ bootstrap samples, which are numerically sufficient and stable based on our experience. 

\begin{table}[!ht]
\centering
\caption{Summary for simulation settings.}\label{table:simulation_setting}
\begin{tabular}{ c c c c c c c}
\hline
  $H_0$ & $\beta_0$ & $\beta_1$ & $\beta_2$ & $\beta^*_0$  & $\beta^*_1$ & $p$\\
\hline
\multicolumn{7}{c}{\underline{type I error settings}}\\
 $\beta_1 = 0$ & -4.23 & 0 & 0.45 & 0.6 & \{0,0.5,1\} & \{0.3,0.5,0.7\}\\
 $\beta_1^* = 0$ & -4.23 & \{0,0.5,1\} & 0.45 & 0.6 & 0 & \{0.3,0.5,0.7\}\\
 $\beta_2 = 0$ & -4.23 & 1 & 0 & 0.6 & \{0,0.5,1\} & \{0.3,0.5,0.7\}\\
\hline
\multicolumn{7}{c}{\underline{power settings}}\\
$\beta_1 = 0$   & -4.23 & \{0,0.2,...,1.8\} & 0.45 & 0.6 & 1 & 0.5\\
 $\beta_1^* = 0$ & -4.23 & 1 & 0.45 & 0.6 & \{0,0.2,...,1.8\} & 0.5\\
 $\beta_2 = 0$   & -4.23 & 1 & \{0,0.2,...,1.8\} & 0.6 & 1 & 0.5\\
\hline
\end{tabular}
\end{table}

\subsection{Hypothesis Testing Results} \label{s:simulation-Hypothesis}
\begin{figure}[!ht]
\begin{subfigure}{0.49\textwidth}
\centering
\includegraphics[width = 0.9\textwidth]{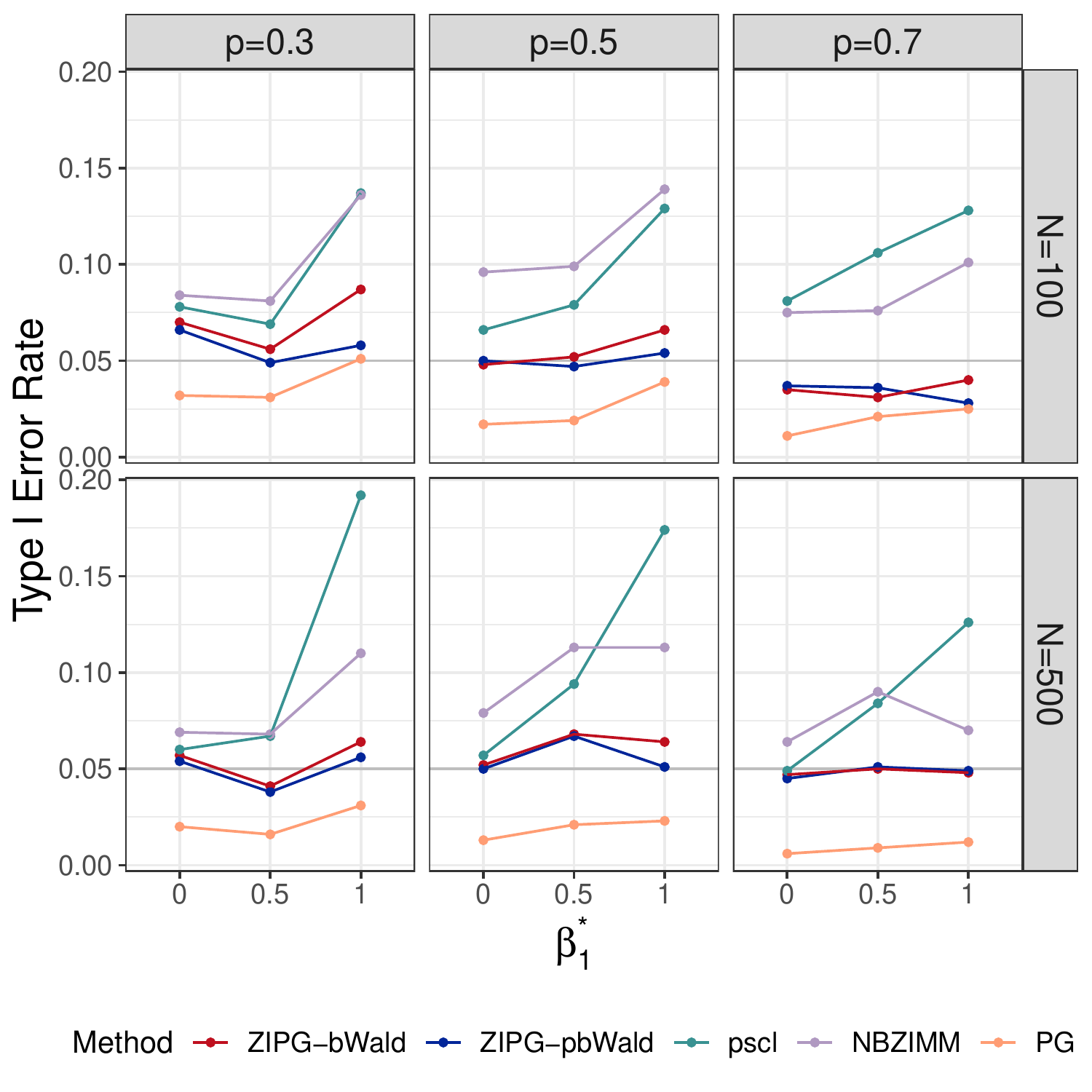}
\caption{Type I error of $H_0: \beta_1 = 0$}\label{fig:type1_beta1}
\end{subfigure}
\begin{subfigure}{0.49\textwidth}
\centering
\includegraphics[width = 0.9\textwidth]{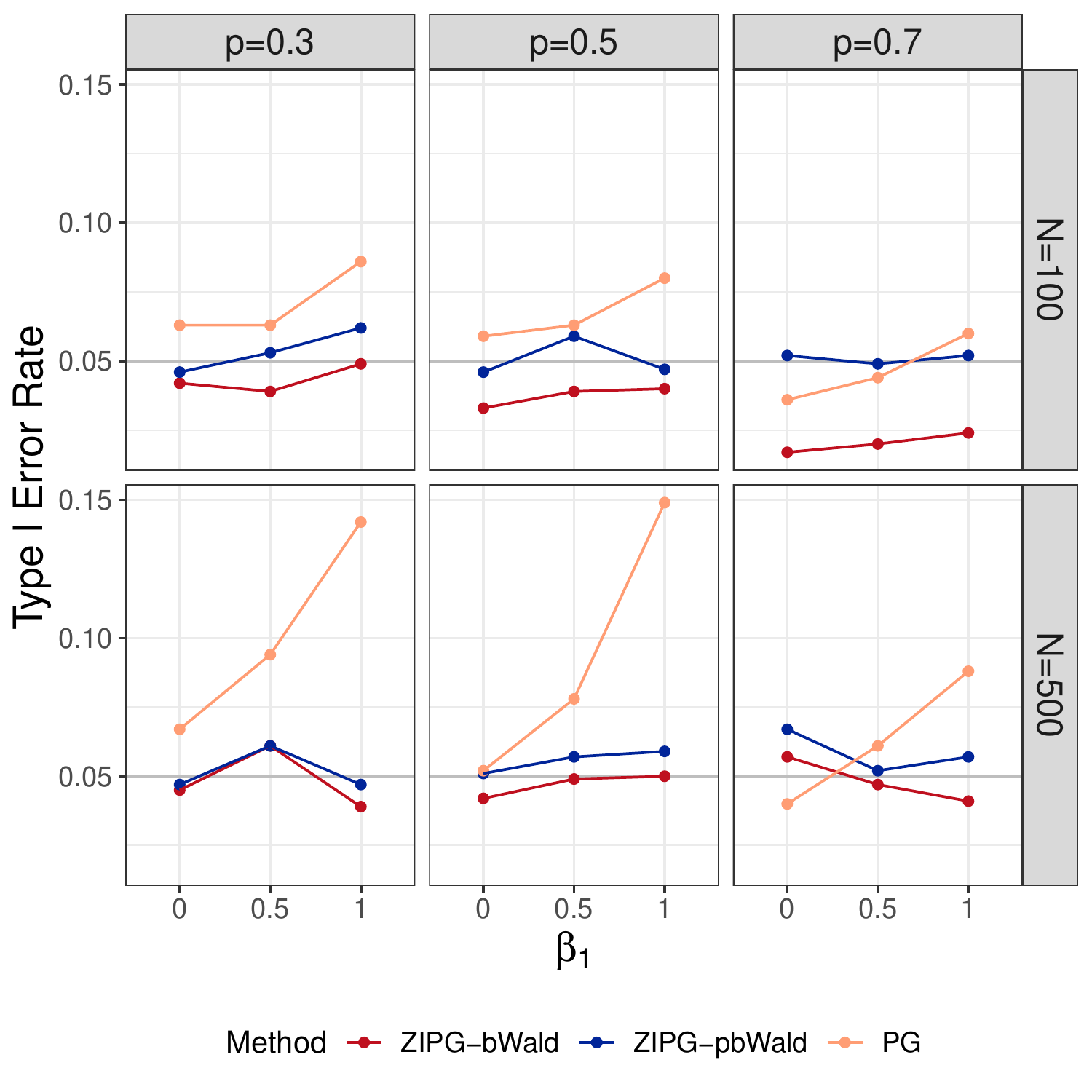}
\caption{Type I error of $H_0: \beta_1^* = 0$}\label{fig:type1_beta1star}
\end{subfigure}
\caption{ Type I error results for (a) $\beta_1$  and (b) $\beta_1^*$. The significance level is $\alpha = 0.05$.} 
\label{fig:type1}
\end{figure}

\paragraph{Type I error analysis for $\beta_1$ and $\beta_1^*$.}
For $H_0: \beta_1 = 0$ in the mean model, we observe that both pscl and NBZIMM have inflated type I errors under all simulation scenarios, and PG is too conservative regardless of the proportion $p$ of inflated zeros (Figure~\ref{fig:type1_beta1}). In particular, the type I error of pscl and NBZIMM increase significantly with the increase of $\beta_1^*$, suggesting that ignoring differential variability could lead to severely increased false positives for the mean model. Of note, the type I error of ZIPG with the bootstrap-based Wald test (ZIPG-bWald) is slightly inflated in a few cases with a small total number of observations ($N=100$), while its results are satisfactory with a larger sample size ($N=500$). In general, the type I error of ZIPG is robustly controlled at the nominal level $\alpha=0.05$, regardless of the change of $\beta_1^*$ and $p$. 
 
For $H_0: \beta_1^* = 0$ in the dispersion model, we observe that ZIPG-bWald maintains a controlled type I error for $\beta^*_1$, yet a little conservative when $N=100$, while ZIPG-pbWald controls type I error to 0.05 more robustly (Figure~\ref{fig:type1_beta1star}). Therefore, we suggest using ZIPG-pbWald as an alternative for hypothesis testing in small-sample scenarios. However, when we have a larger sample size ($N=500$), both ZIPG-bWald and ZIPG-pbWald perform similarly. 
 
\begin{figure}[!ht]
\begin{subfigure}{0.49\textwidth}
\centering
\includegraphics[width = 0.9\textwidth]{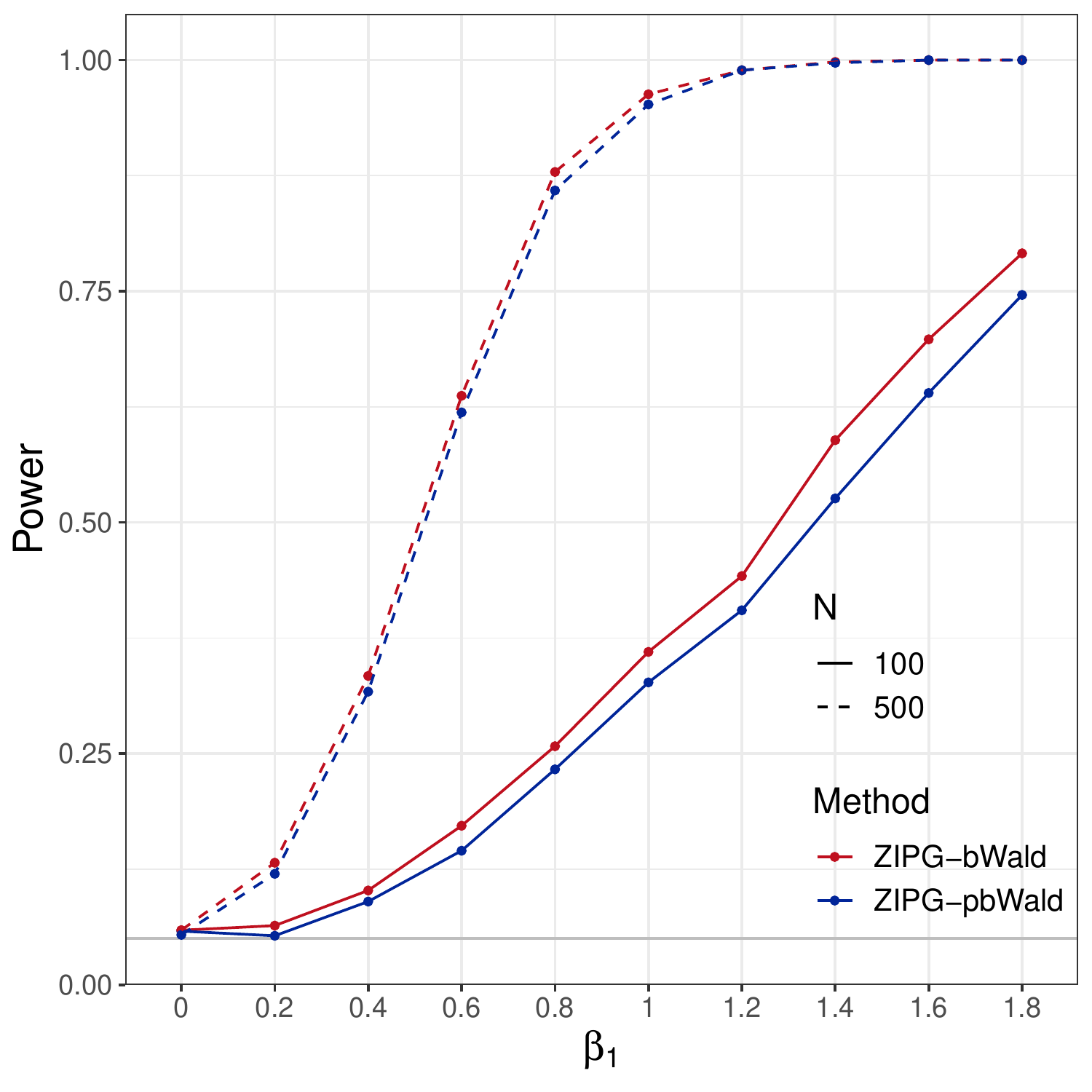}
\caption{Power of rejecting $H_0: \beta_1 = 0$}\label{fig:power_beta1}
\end{subfigure}
\begin{subfigure}{0.49\textwidth}
\centering
\includegraphics[width = 0.9\textwidth]{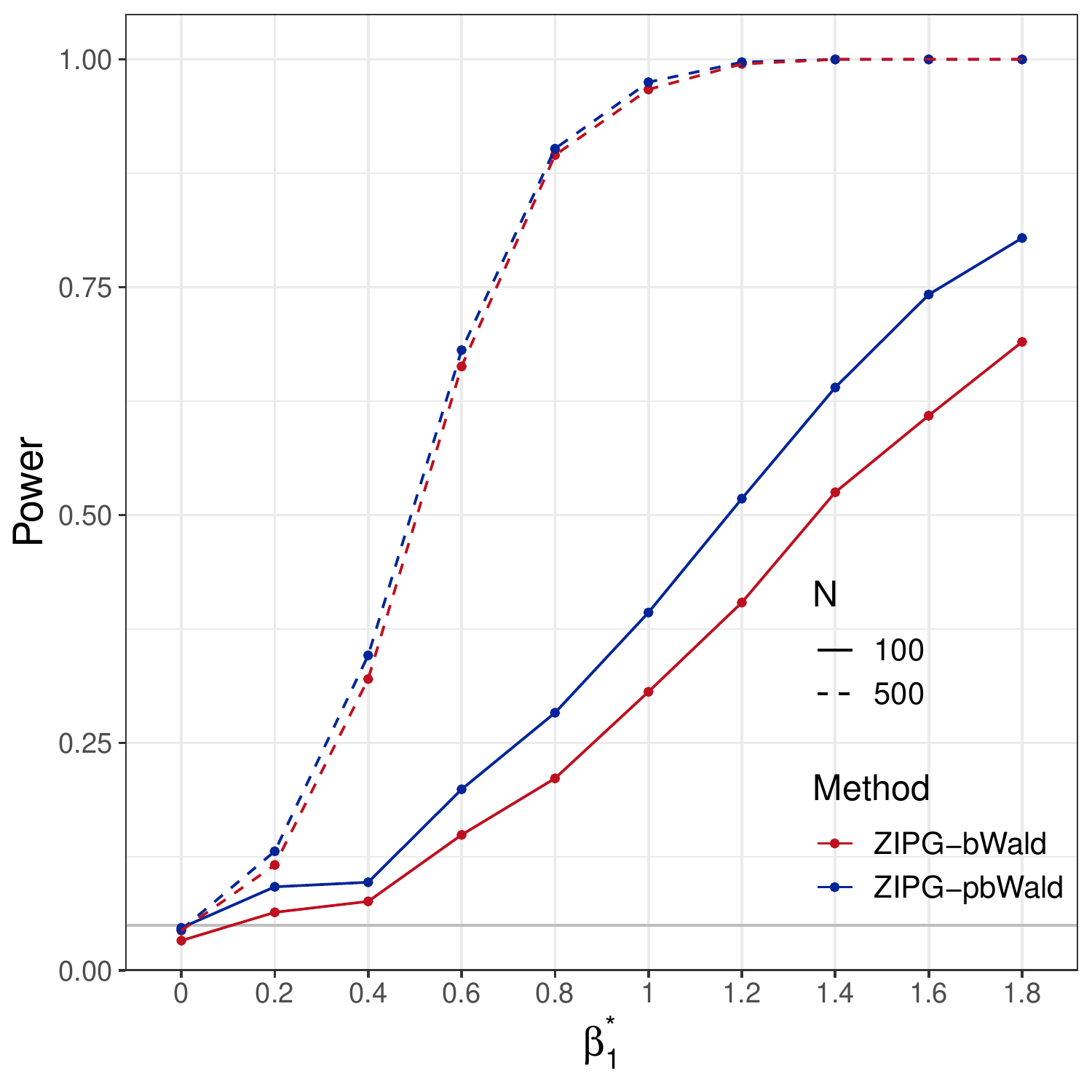}
\caption{Power of rejecting $H_0: \beta_1^* = 0$}\label{fig:power_beta1star}
\end{subfigure}
\caption{Power curves of rejecting null hypothesis with $N = 100$ (solid lines) and $N = 500$ (dash lines). With $p =0.5$, the proportion of observed zeros decreased from 0.606 to 0.582 with the increase of $\beta_1$ in (a), while it increased from 0.536 to 0.653 with the increase of $\beta_1^*$ in (b).} 
\label{fig:power}
\end{figure}

\paragraph{Power analysis for $\beta_1$ and $\beta_1^*$.}
We present power results of testing $H_0: \beta_1 = 0$ (Figure~\ref{fig:power_beta1}) and $H_0: \beta_1^* = 0$ (Figure~\ref{fig:power_beta1star}). Since pscl, NBZIMM, and PG fail to preserve the nominal type I error, we do not evaluate their empirical power in the power analysis. For both null hypotheses, the power curves increase with the increase of sample size and true effect size. ZIPG-pbWald and ZIPG-bWald perform similarly in larger sample cases, while ZIPG-pbWald is relatively more powerful for detecting differential variability ($H_0: \beta_1^* = 0$) with a small sample size ($N=100$).
We only report the common covariates in both the mean model and dispersion model in the main text, and hypothesis testing results on covariates only in the mean model are reported in Section 3.3 of the Supplements.

\subsection{Point and Interval Estimation Results}
To demonstrate the advantage of ZIPG regarding point estimators and confidence intervals,
we report the average bias (e.g., $\{\sum_l(\hat\beta_l - \beta)\}/L$) with its standard error over $l=1,\ldots,L=1000$ Monte Carlo replicates, average bootstrap standard error (avg-SE, e.g., $\{\sum_l\widehat{SE}(\hat\beta_l)\}/L$), root mean squared error (RMSE, e.g., $\sqrt{\{\sum_l(\hat\beta_l - \beta)^2)\}/L}$), and converge rate of confidence intervals (CR) for the settings with $\beta_1 = 0, \beta^*_1 = 1$ (Figure~\ref{fig:type1_beta1}) and $\beta_1 = 1, \beta^*_1 = 0$ (Figure~\ref{fig:type1_beta1star}) with $p=\{0.5,0.7\}$ and $N=500$. Results under other settings are similar and hence not reported.

In Table \ref{tab:type1}, we observe that ZIPG-bWald has the smallest bias of $\beta_1$, $\beta_1^*$ and $\gamma$ among all methods when $\beta_1^* = 1$. In addition, ZIPG is often more efficient than the other two methods, providing a smaller RMSE. For $\beta_1$ and $\beta_1^*$, ZIPG-bWald always maintains a valid confidence interval with its coverage rate close to nominal level 0.95, whereas pscl and NBZIMM provide underestimated confidence intervals in most cases, and PG estimate $\beta_1^*$ with strong bias and provide a more conservative CI for $\beta_1$. Additional results with setting $\beta_1 = 1, \beta^*_1 = 0$ corresponds to Figure~\ref{fig:type1_beta1star} are presented in Section 3.4 of the Supplements.

\begin{table}[!ht]\footnotesize
\caption{Average bias and its SE, average standard error, RMSE, and the empirical coverage rate (CR) of $\beta_1$ and $\beta_1^*$ estimators. Simulation parameters are set as $\beta_1 = 0, \beta^*_1 = 1$ corresponding to Figure~\ref{fig:type1_beta1};  $p = \{0.5,0.7\}$ or $\gamma = \{0,0.847\}$. NBZIMM does not report inference and CI on $\gamma$. The proportion of zeros observed in each simulation setting is denoted as $p_{\rm obs}$. }\label{tab:type1}
\begin{minipage}{0.49\linewidth}
\centering
\setlength{\tabcolsep}{0.4mm}{
\begin{tabular}{l l l l l l}
\hline
 & Method & avg-bias(SE) & avg-SE & RMSE & CR\\
\hline
\multicolumn{6}{c}{$N=500$, $\beta_1 = 0$, $\beta_1^* = 1$, $p= 0.5$, $p_{\rm obs} = 0.61$}\\ 
\hline 
$\beta_1$ & \textbf{ZIPG-bWald} & -0.013(0.254) & 0.254 & 0.254 & 0.936 \\ 
 & pscl & 0.216(0.252) & 0.23 & 0.332 & 0.826 \\ 
 & NBZIMM & 0.084(0.274) & 0.235 & 0.286 & 0.877 \\ 
 & PG & -0.016(0.257) & 0.307 & 0.258 & 0.977 \\ 
$\beta_1^*$ & \textbf{ZIPG-bWald} & -0.009(0.246) & 0.256 & 0.246 & 0.954 \\ 
 & PG & -0.449(0.169) & 0.165 & 0.48 & 0.231 \\ 
$\gamma$ & \textbf{ZIPG-bWald} & -0.004(0.145) & 0.158 & 0.145 & 0.968 \\ 
 & pscl & 0.108(0.152) & 0.154 & 0.186 & 0.828 \\ 
 & NBZIMM & -0.245(0.454) & - & 0.516 & - \\ 
\hline 
\end{tabular}}
\end{minipage}
\begin{minipage}{0.49\linewidth}  
\centering
\setlength{\tabcolsep}{0.6mm}{
\begin{tabular}{l l l l l l}
\hline
 & Method & avg-bias(SE) & avg-SE & RMSE & CR \\
\hline
\multicolumn{6}{c}{$N=500$, $\beta_1 = 0$, $\beta_1^* = 1$, $p= 0.7$, $p_{\rm obs} = 0.77$}\\ 
\hline 
$\beta_1$ & \textbf{ZIPG-bWald} & -0.027(0.326) & 0.341 & 0.327 & 0.952 \\ 
 & pscl & 0.220(0.311) & 0.299 & 0.381 & 0.874 \\ 
 & NBZIMM & -0.07(0.408) & 0.363 & 0.414 & 0.917 \\ 
 & PG & -0.037(0.34) & 0.432 & 0.342 & 0.988 \\ 
$\beta_1^*$ & \textbf{ZIPG-bWald} & 0.006(0.341) & 0.373 & 0.341 & 0.959 \\ 
 & PG & -0.536(0.21) & 0.211 & 0.576 & 0.282 \\ 
$\gamma$ & \textbf{ZIPG-bWald} & 0.007(0.158) & 0.165 & 0.158 & 0.958 \\ 
 & pscl & 0.047(0.872) & 0.153 & 0.873 & 0.864 \\ 
 & NBZIMM & -0.294(0.604) & - & 0.671 & - \\ 
\hline 
\end{tabular}
 }
\end{minipage}
\end{table}

\subsection{Model Sensitivity Analysis}\label{s:Model-sensitivity}
While ZIPG assumes that all differential variability comes from the Poisson-Gamma part (i.e., $\theta$) and the true zero proportion $p$ is only taxon-specific, one may wonder how the model fits when these assumptions are violated. Here, we evaluate ZIPG under the misspecified model, in which the true zero proportion $p$ is also associated with covariates (denoted as ``ZIPG-full"). Of note, ZIPG-full is equivalent to Omnibus \citep{chen2018omnibus} from a hypothesis testing perspective, whereas the Omnibus test does not provide point/interval estimation and hypothesis test for each parameter separately. We consider the group covariates $X_1$ only and sample size $N=500$. We use a logistic link ${\rm logit}(p) = \gamma_0 + \gamma_1 X_1$ with $(\beta_0,\beta_0^*,\gamma_0) = (-4.23,0.6,-0.847)$, and then we report the performance of ZIPG with  $\gamma_1$ increasing from 0 to 2.5, which is equivalent to increasing $p$ from $0.30$ to $0.84$ in the group with $X_1 = 1$. 

\begin{figure}[!h]
\centering
\includegraphics[width = 0.6\textwidth]{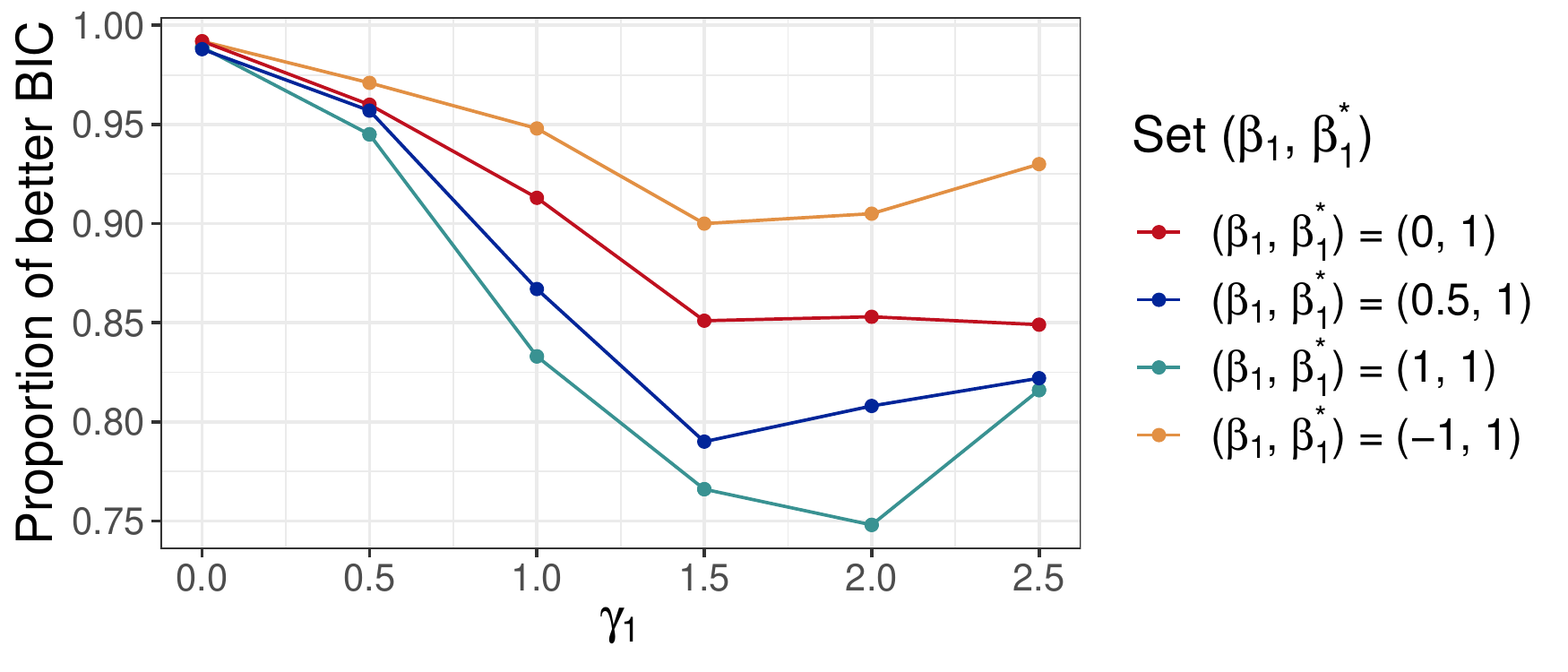}
\caption{
Proportion of ZIPG having smaller BIC than ZIPG-full in each simulation setting when $N = 500$.}
\label{fig:gamma}
\end{figure}

For two model fittings, i.e., ZIPG and ZIPG-full, we report the proportion of simulations suggesting better BIC from ZIPG (Figure~\ref{fig:gamma}). Over 1000 replicated simulations in each setting, ZIPG has a smaller BIC than ZIPG-full in most cases. Moreover, we also use Kolmogorov-Smirnov test \citep{kolmogorov1933sulla,smirnoff1939ecarts} to  compare the ZIPG predicted distribution with the simulated observations: 100\% replicates report  insignificant differences between the two distributions ($p>0.05$)
, suggesting that ZIPG-predicted distribution has no difference to the observed samples. 

We also evaluate ZIPG's performance under other misspecified models: (1) Poisson-Gamma without zero inflation and (2) zero-inflated Beta-Binomial model (see Section 3.5 of the Supplements). In both scenarios, ZIPG preserves nominal type I error and retains its superior power in detecting differential abundance/variability, especially for large-sample cases.

\subsection{Additional Simulation Results}\label{sec:sim_additional}

 {\color{black}
 We conduct additional simulations to compare multiple ways of constructing test statistics and confidence intervals.
 For hypothesis testing,  we investigate different test statistics for the proposed ZIPG (Section 3.1 of the Supplements), including the Wald test without bootstrap (ZIPG-Wald) and the likelihood ratio test (ZIPG-LRT). Simulation results suggest that ZIPG with bootstrap-based Wald tests (i.e., ZIPG-bWald and ZIPG-pbWald) are desired. For confidence intervals, we compare the coverage rate for both nonparametric and parametric bootstrap through different construction strategies, such as normality-based/quantile-based/$\rm BC_a$\citep{efron1994introduction} intervals. Results show that the normality-based confidence interval often has close-to-nominal coverage with a low computational cost (Section 3.2 of the Supplements).  
 
 We further evaluate different resampling schemes for ZIPG-bWald, such as resampling based on measurements or subjects. Given the total sample size $N=200$, results suggest no significant difference between the two strategies (Section 3.6 of the Supplements).
 
 Additional simulations about how measurement times affect ZIPG performance are provided in Section~3.6 of the Supplements. We consider the following two scenarios: (1) when the numbers of measurements per subject are very small (i.e., $m=2$) and (2) when the numbers of measurements per subject are unequal. 
 Results show that the proposed ZIPG method is valid for both cases above. }

\section{Data Analysis} \label{s:application} 
In this section, we analyze two microbiome data sets, Romero \citep{romero2014composition} and Dietary \citep{johnson2019daily}, to investigate how physical conditions impact microbiome stability in specific taxa. The proposed ZIPG method with bootstrap-based Wald test is compared to pscl \citep{Rpscl_zeroinfl} and DESeq2 \citep{DESeq2014} from two perspectives: identification of taxa related to covariates and goodness of fit for prediction models. In addition, we also report hypothesis testing results from Omnibus \citep{chen2018omnibus} to validate the results of ZIPG. We also performed NBZIMM, but it detected only a few taxa with their default subject-level random effect, {\color{black} and hence we present the corresponding results in Section~4.3 of the Supplements.}

\subsection{Data Description}
\paragraph{Romero} is a longitudinal case-control study including 16s rRNA gene sequence-based vaginal microbiota from 22 pregnant and 32 non-pregnant women with samples collected from each subject over intervals of weeks, resulting in 143 taxa and $N=900$ longitudinal samples ({\color{black} 139 measurements from pregnant women and 761 measurements from non-pregnant women.}) To investigate how taxa are impacted by pregnant status and age in this data, we set the covariates matrix $\boldsymbol{X^*}=\boldsymbol{X} = (X_1\ X_2\ X_3)$, where $X_1$ is a binary indicator of pregnant status, $X_2$ is the observational age, and $X_3$ is an indicator for the race (white or others) but not of our main interest. Note that both pregnant status and age are not changed for each person during data collection.

\paragraph{Dietary} is diet-microbiome data with shotgun metagenomic sequencing results of fecal samples and daily dietary records of 34 subjects on 17 consecutive days. There are total $N=475$ samples with both microbiome data and dietary records available. In this data, the main analysis of interest is how alcohol affects the microbiome variability. We created a binary indicator for {\color{black} 25 alcohol drinkers and 9 teetotalers}, and included this variable in both $\boldsymbol{X^*}$ and $\boldsymbol{X}$. To account for the impact of other dietary intakes, we also include the first two principal components of the macronutrient matrix in $\boldsymbol{X}$. 

In microbiome studies, it is common to filter out taxa with extremely low abundance (i.e., $p_{\rm obs}>0.9$) for more stable and reliable results \citep{wadsworth2017integrativea,jiang2021bayesian,zhang2020fast}. Further, taxa with $p_{\rm obs}<0.1$ are likely to have little or no zero inflation and can be modeled by other existing methods. Though ZIPG can still be performed with satisfactory results (see Section 3.5 of the Supplements), this group of taxa is not of our main interest. For both Romero and Dietary data, we analyze the taxa with $0.1<p_{\rm obs}<0.9$, which results in 25 taxa in Romero and 52 taxa in Dietary. More details about $p_{\rm obs}$ in Romero and Dietary can be found in Section 4.1 of the Supplements. To account for multiple testing, we report the results with the controlled false discovery rate (FDR$<0.05$) using the method of \cite{benjamini1995controlling}.

\subsection{Results on Hypothesis Testing}
\begin{figure}[!ht]
\begin{subfigure}{0.32\textwidth}
  \centering
  \includegraphics[width = 0.9\textwidth]{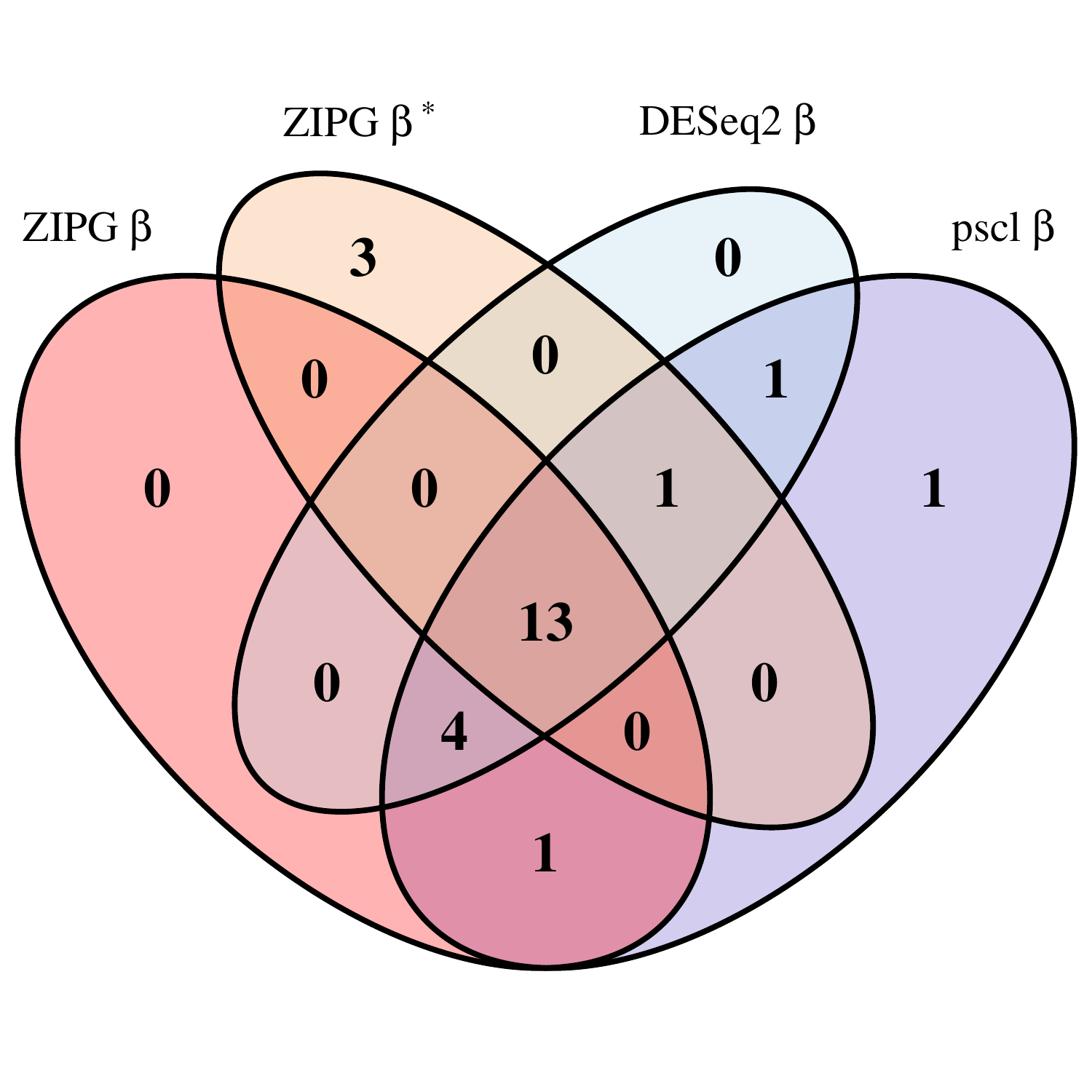}
  \caption{Romero Pregnant}
  \label{fig:Venn-Pre}
\end{subfigure}
\begin{subfigure}{.32\textwidth}
  \centering
  \includegraphics[width = 0.9\textwidth]{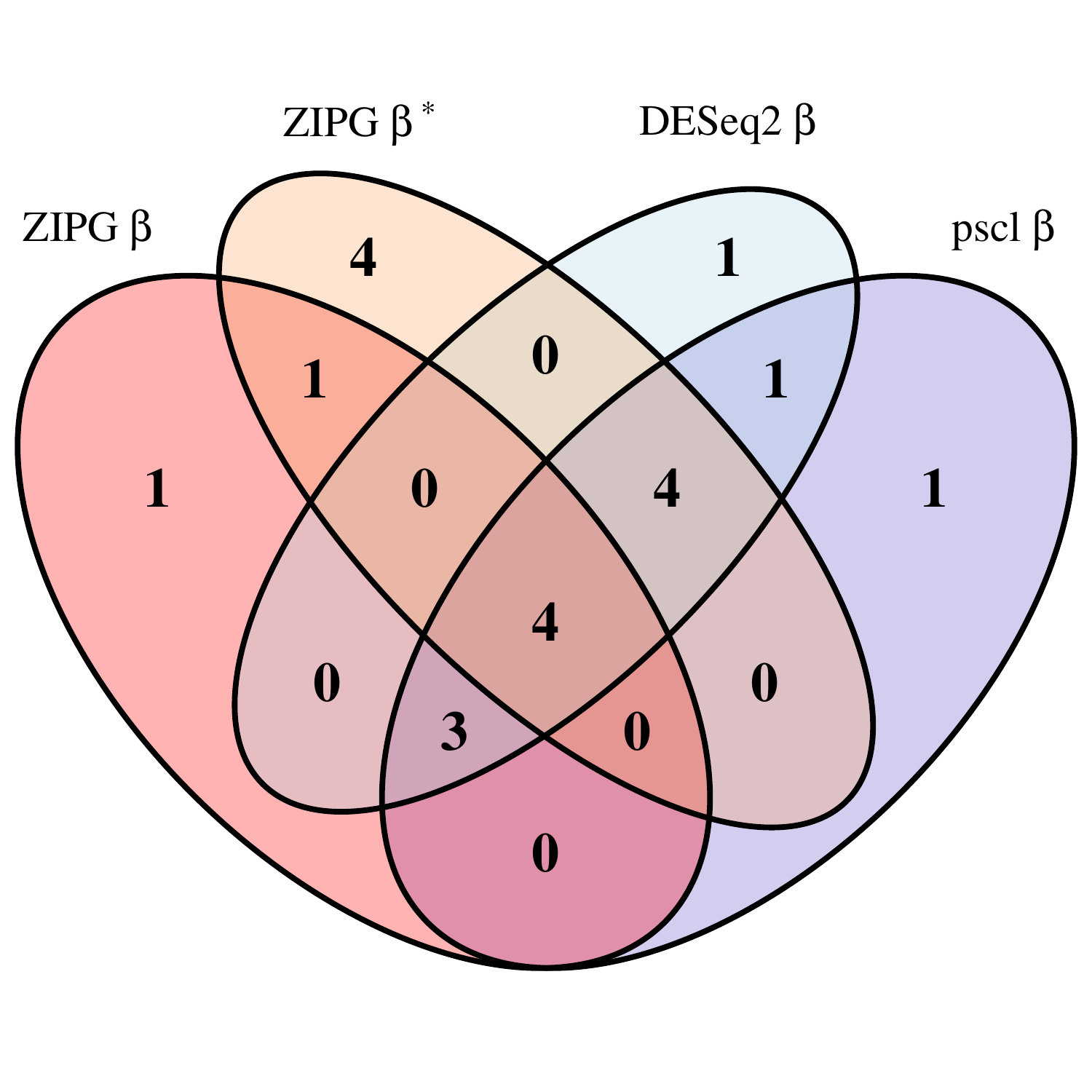}
  \caption{Romero Age}
  \label{fig:Venn-Age}
\end{subfigure}
\begin{subfigure}{.32\textwidth}
  \centering
  \includegraphics[width = 0.9\textwidth]{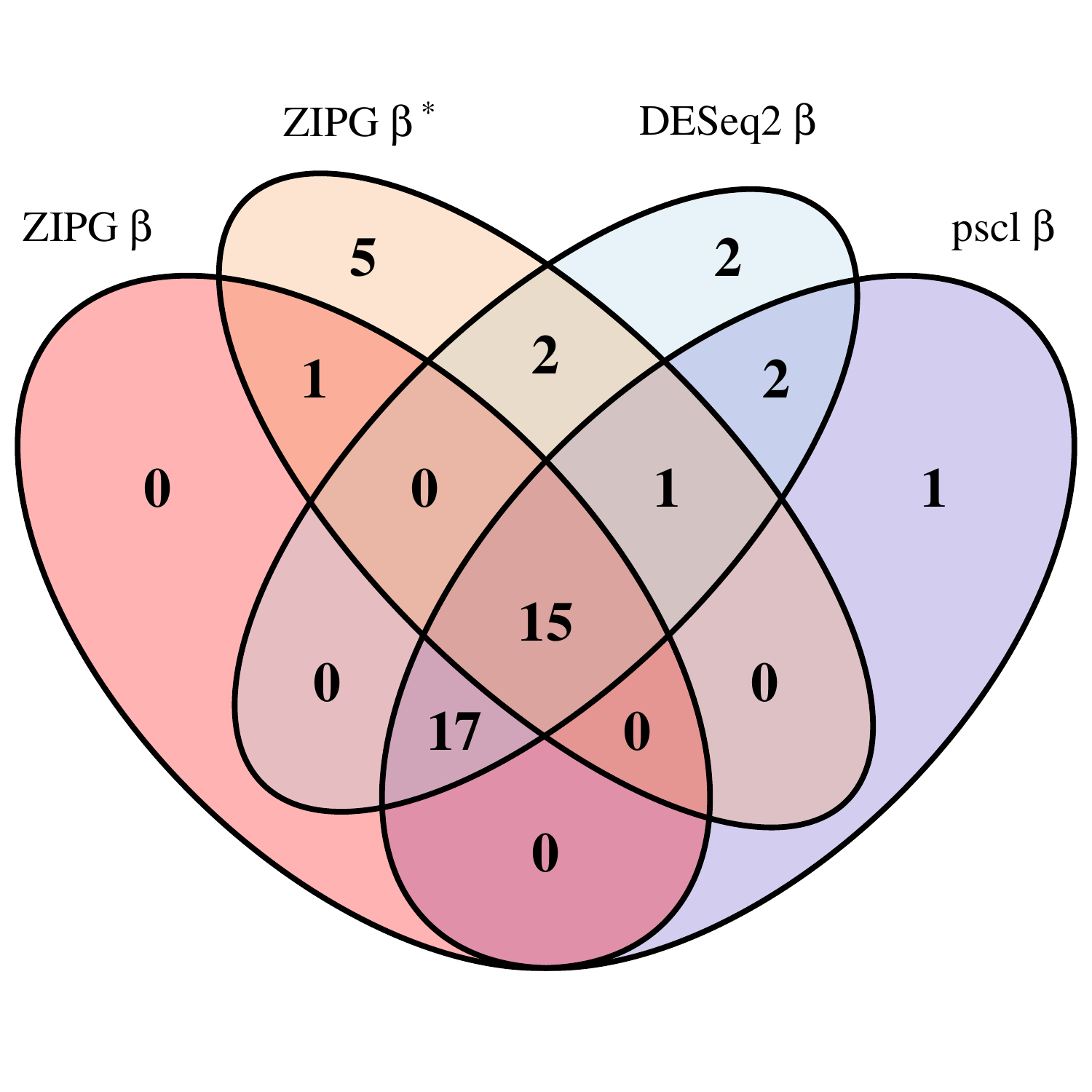} 
  \caption{Dietary ALC}
  \label{fig:Venn-ALC}
\end{subfigure}
\caption{The numbers of taxa with significant difference detected by ZIPG, DESeq2 and pscl after controlling FDR$<0.05$ regarding $H_0 : \beta= 0$ and $H_0 : \beta^* = 0$ in Romero (\ref{fig:Venn-Pre} and \ref{fig:Venn-Age}) and Dietary (\ref{fig:Venn-ALC}). }
\label{fig:Venn}
\end{figure}

We first present numbers of identified taxa regarding the covariates of interest in the two studies (Figure~\ref{fig:Venn}). For ZIPG, we show the set of taxa associated with $\boldsymbol{X}$ in the mean model (denoted as ZIPG $\beta$) and  $\boldsymbol{X^*}$ in the dispersion model (denoted as ZIPG $\beta^*$), separately. In Romero, we observe that pregnant subjects are clustered under age 35, while non-pregnant subjects are collected in a much wider range of ages. Owing to the unbalanced sample {\color{black}that pregnant women have fewer measurements and are often younger than non-pregnant women}, we use parametric bootstrap (i.e., ZIPG-pbWald) for more stable results. ZIPG identified 18 taxa with differential abundance ($H_0 : \beta_1 = 0$) and 17 taxa with differential variability ($H_0 : \beta_1^* = 0$) associated with pregnancy, while
13 taxa are associated with pregnant status with both abundance and variability (Figure \ref{fig:Venn-Pre}). Most of the taxa found by pscl and DESeq2 are also identified by ZIPG, while ZIPG also identified 3 additional taxa with significant differential variability, which are not detected by other methods. {\color{black}Further, ZIPG identified totally additionally 6 taxa associated to age with differential abundance ($H_0 : \beta_2 = 0$) or differential variability ($H_0 : \beta_2^* = 0$) , compared to other methods (Figure \ref{fig:Venn-Age})}. 

In Dietary, ZIPG is performed using nonparametric bootstrap Wald test (i.e., ZIPG-bWald), because the data is balanced and the sample size is sufficient. ZIPG identified 33 taxa with differential abundance ($H_0 : \beta_1 = 0$) and 24 taxa with differential variability ($H_0 : \beta_1^* = 0$) associated to the alcohol intake (Figure \ref{fig:Venn-ALC}). Compared to other methods, ZIPG discovered 5 extra taxa only associated with differential variability and 1 extra taxon associated with both differential abundance and differential variability.

We further use the Omnibus test \citep{chen2018omnibus} as verification for ZIPG detected taxa. The Omnibus test links covariates of interest to all three parameters in the Negative-Binomial distribution and rejects the null hypothesis if any covariate is associated with any of the parameters. Thus,  taxa identified by both ZIPG and Omnibus are less likely to be false positives. As expected, in Romero, all taxa identified by ZIPG are also detected by Omnibus. In Dietary, 35 out of 41 taxa identified by ZIPG are also detected by Omnibus. Though the Omnibus test detected more taxa than ZIPG, it is worth pointing out that the Omnibus test can not distinguish differential abundance and differential variability and does not provide point/interval estimation as ZIPG does. Details of taxa detected by each method and estimation results for those taxa regarding parameters of interest are shown in Section 4.3 of the Supplements.

\subsection{Analysis on Model Fitting}
To visualize the differential variability tested by ZIPG (i.e., $H_0 : \beta^* = 0$),  we further analyze the result of Bifidobacteriaceae and Lactobacillus.vaginalis from Romero, and  Burkholderiales bacterium and Alistipes indistinctus from Dietary as examples. {\color{black}Other taxa with differential variability identified by ZIPG have similar conclusions.} 

\begin{figure}[!ht]
\centering
\begin{subfigure}{0.588\textwidth}
  \centering
\includegraphics[width = 0.93\textwidth]{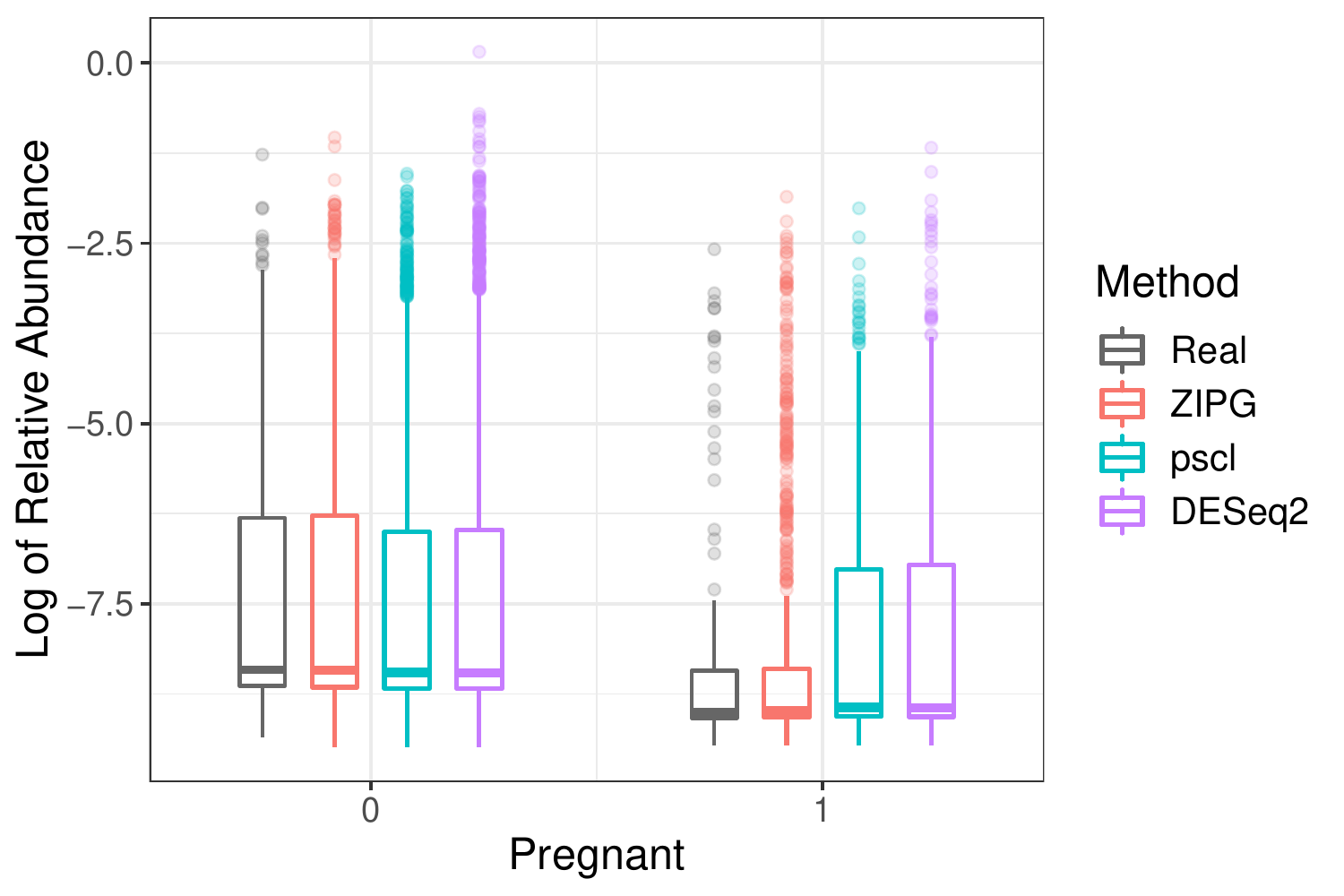}
  \caption{\color{black}Box plot of predicted distribution}
  \label{fig:Romero-boxplot}
\end{subfigure}
\begin{subfigure}{0.392\textwidth}
  \centering
\includegraphics[width = 0.93\textwidth]{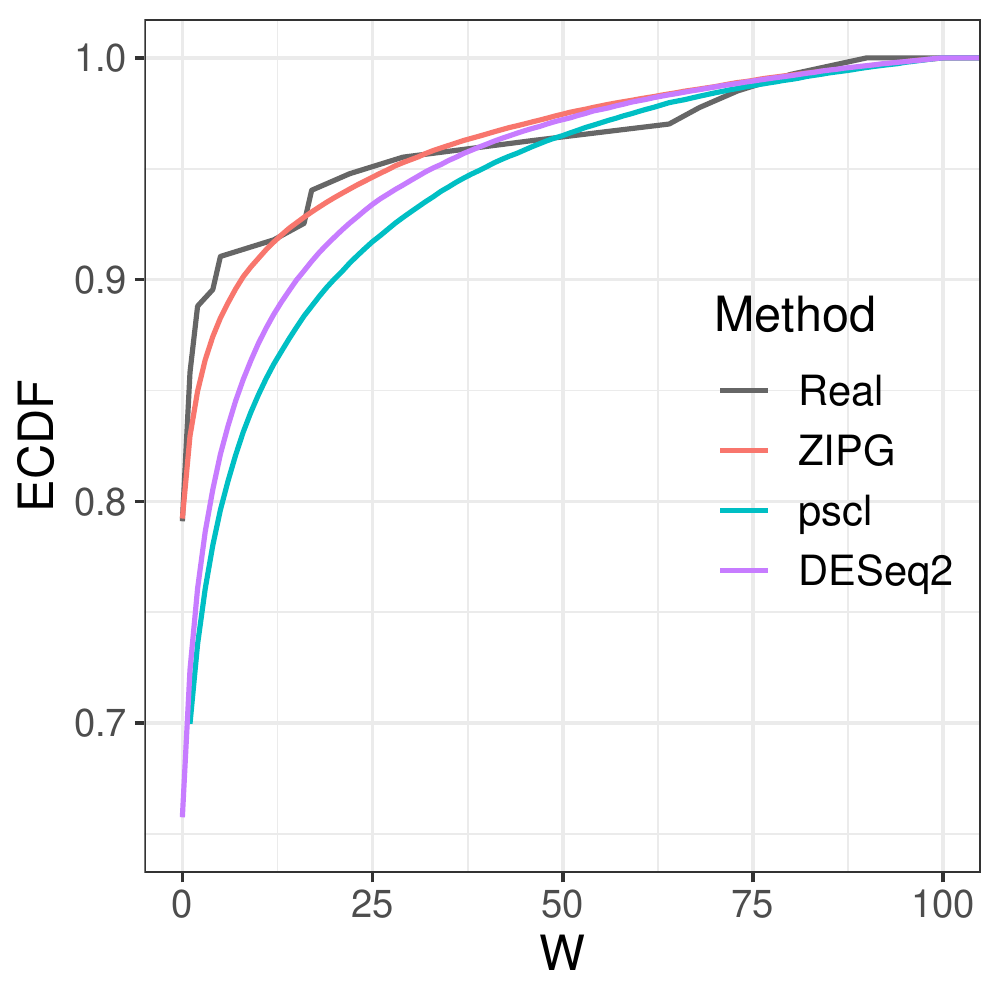}
  \caption{ECDF (Pregnant)}
  \label{fig:Romero-ECDF}
\end{subfigure}
\caption{Bifidobacteriaceae in Romero: (a) box plot for the predicted distribution and the real observed counts (the log of relative abundance is presented with zero count samples adjusted to 0.5 in the pregnant and non-pregnant group), (b) ECDF in the pregnant group. 
}
\label{fig:Example_Romero_taxa17}
\end{figure}

First, we compare the goodness of fit of results from fitted models to the empirical distribution of the relative abundance.  The log of relative abundance observed in real data (i.e., Bifidobacteriaceae from Romero) is compared to the predicted distribution according to the estimated model by ZIPG, pscl, and DESeq2 (Figure \ref{fig:Example_Romero_taxa17}). {\color{black}For boxplots, we generated samples from each predicted distribution with a quintuple of the observed sample sizes for a better visualization at the tail (e.g., Figure \ref{fig:Example_Romero_taxa17}(\subref{fig:Romero-boxplot})).} All three methods can estimate the median of two groups (pregnant and non-pregnant) accurately, but pscl and DESeq2 cannot distinguish the overdispersion between the two groups, as the box length for the pregnant and non-pregnant groups are similar.  On the contrary, ZIPG identified the differential variability of this taxon associated with the factor pregnant with $p = 0.00256$ regarding the null hypothesis $H_0 : \beta_1^* = 0$, and thus its simulated data matches the real data better, providing a shorter interquartile range with a long tail in the pregnant group. We also present the empirical cumulative distribution functions (ECDF) of the log of the relative abundance for the pregnant group, using sampled data from fitted ZIPG, pscl, and DESeq2 models (Figure \ref{fig:Example_Romero_taxa17}(\subref{fig:Romero-ECDF})). It has been shown that ZIPG also fits the real data better than other methods. {\color{black}Quantile-quantile plots for real data versus predicted distribution also show that ZIPG models the entire distribution better (Section 4.2 of the Supplements).} 

\begin{figure}[!ht]
\centering
\begin{subfigure}{0.63\textwidth}
  \centering
  \includegraphics[width = 1\textwidth]{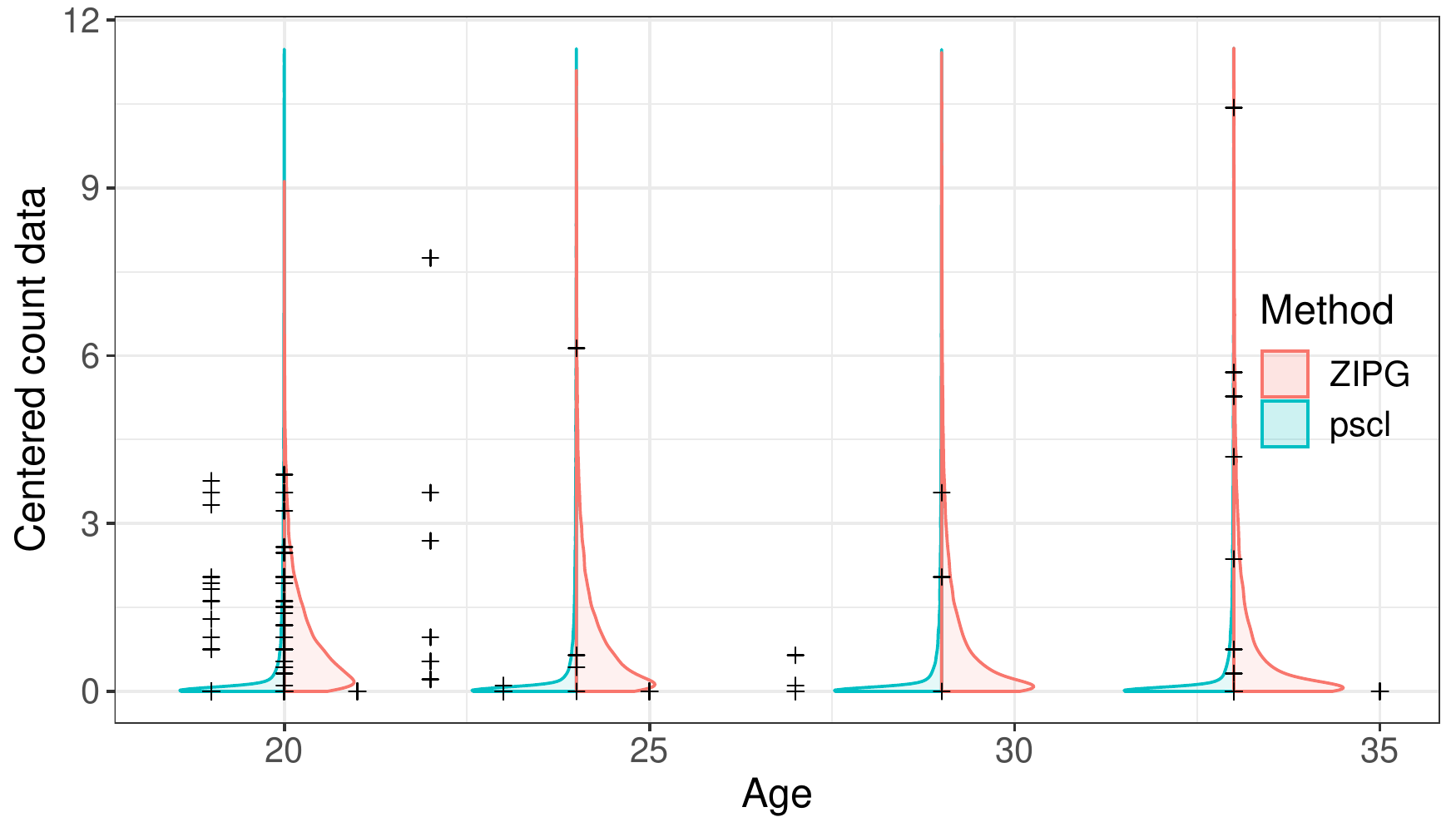}
  \caption{Predicted distribution (Pregnant, White)}
  \label{fig:Romero-Violin32}
\end{subfigure}
\begin{subfigure}{0.36\textwidth}
  \centering
  \includegraphics[width = 1\textwidth]{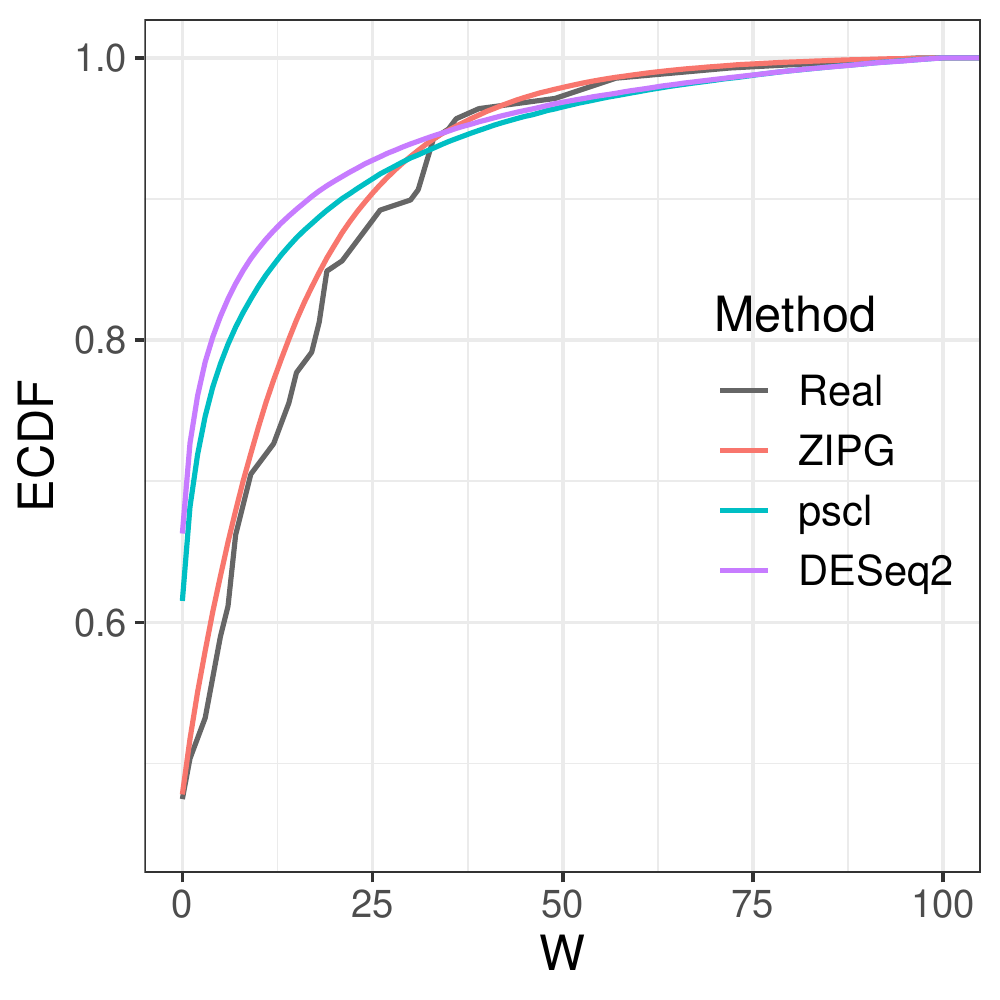}
  \caption{ECDF (Pregnant)}
  \label{fig:Romero-ECDF32}
\end{subfigure}
\caption{
Lactobacillus vaginalis in Romero: (a) the half-violin curves of the fluctuation factor $U$ for the pregnant, white women generating from parameters estimated by ZIPG and pscl respectively, comparing to the raw count data divided by its mean (``+"). (b) ECDF of predicted and real observed counts in the pregnant group.}
\label{fig:Example_Romero_taxa32}
\end{figure}

For Lactobacillus.vaginalis in Romero, we present the relative abundance based on the simulated distributions of the fluctuation factor $U \sim {\rm Gamma}(\theta_{i}^{-1},\theta_{i})$ from ZIPG and pscl at four representative ages (Figure~\ref{fig:Example_Romero_taxa32}). ZIPG identified that the differential variability of this taxon is associated to age with $p$ < 0.001 regarding $H_0 : \beta_2^* = 0$. Accordingly, we observe that the shape of the fitted distribution changes with the increase of age. However, pscl is not able to model the change in the entire shape of distribution as the average relative abundance remained the same in this group. The ECDF plot of the pregnant group again shows that ZIPG fitted model is closer to the empirical distribution.

\begin{figure}[!ht]
 \centering
\begin{subfigure}{0.49\textwidth}
  \centering
\includegraphics[width = 0.99\textwidth]{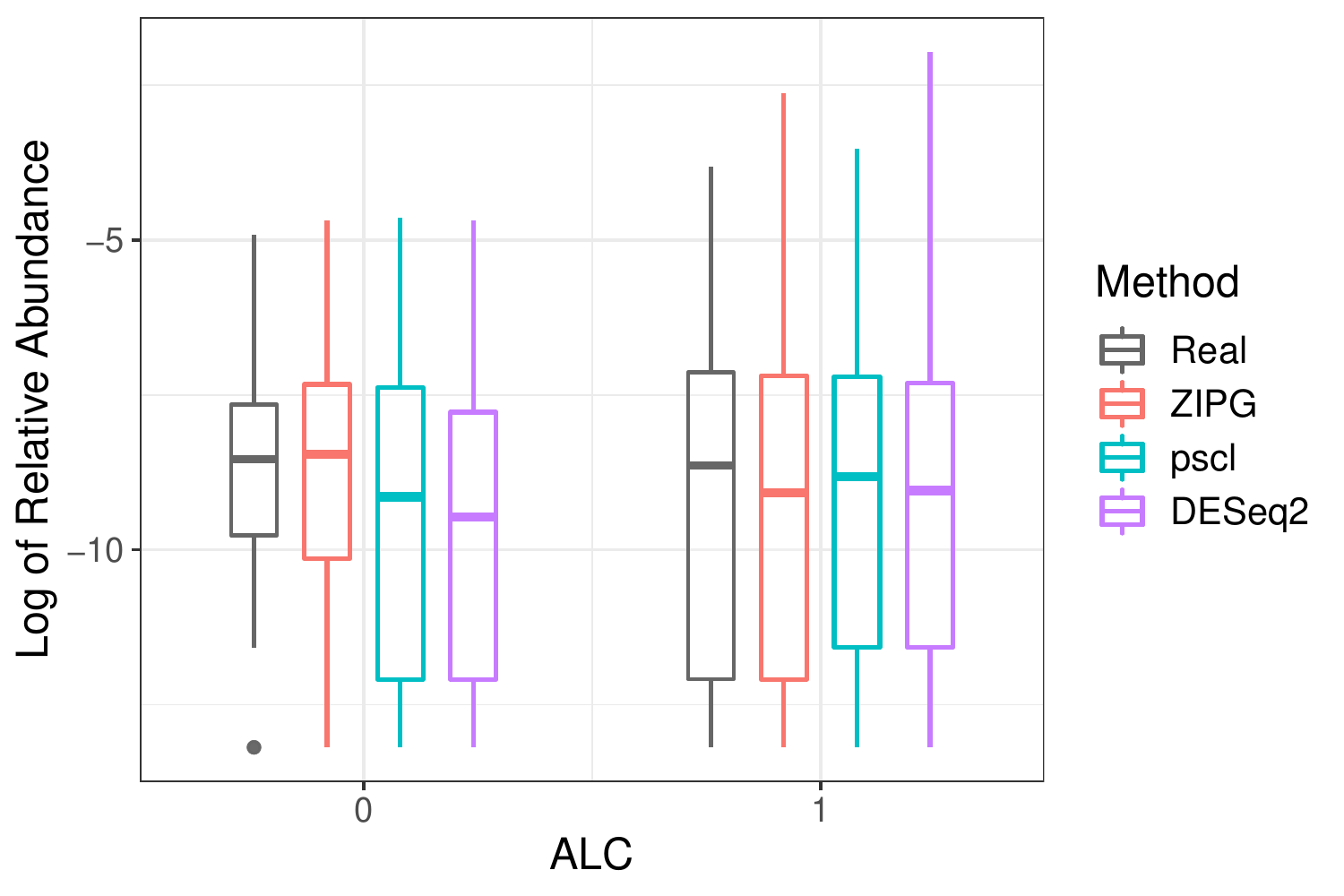}
  \caption{Burkholderiales bacterium }
\end{subfigure}
\begin{subfigure}{0.49\textwidth}
  \centering
\includegraphics[width = 0.99\textwidth]{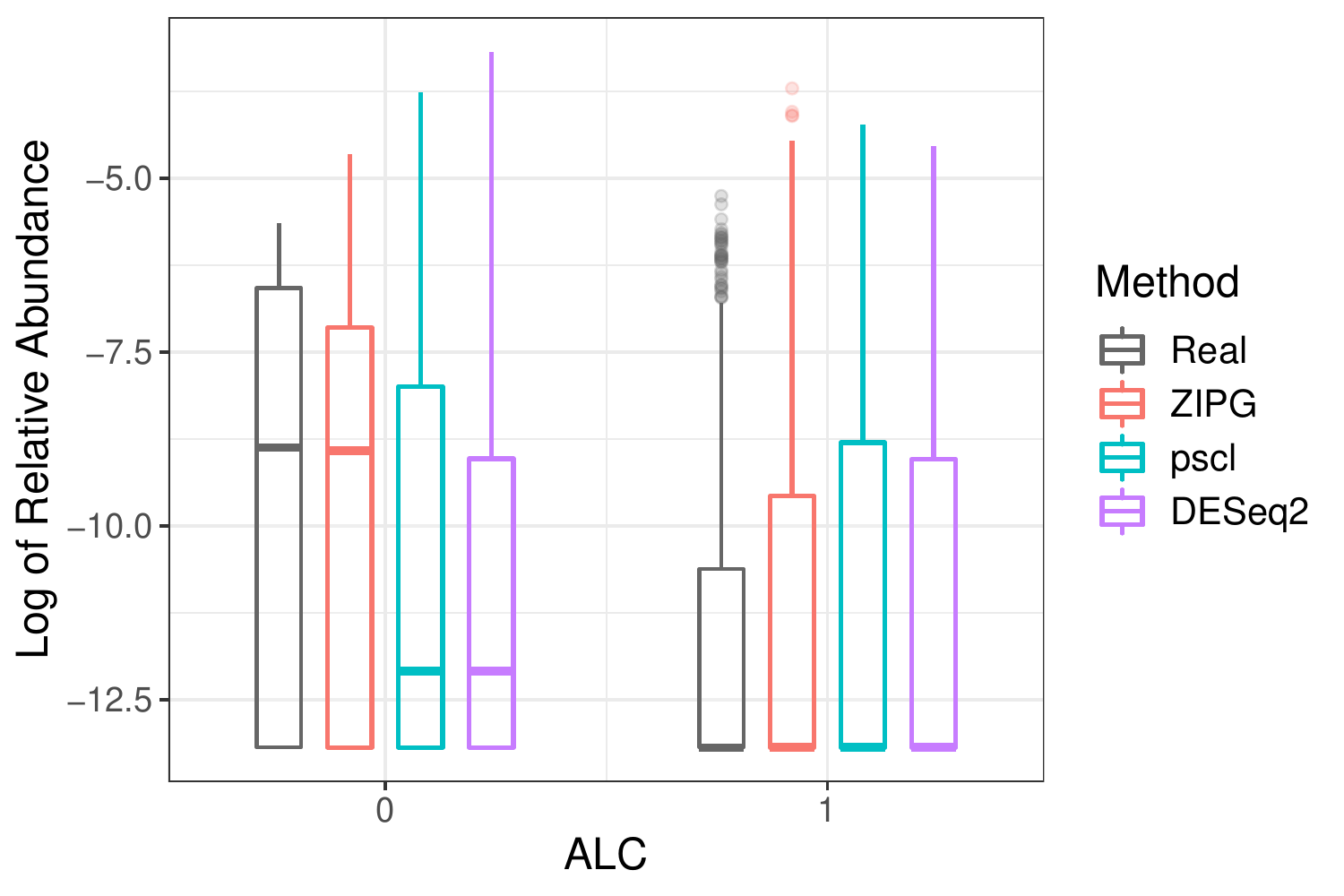}
  \caption{Alistipes indistinctus}
\end{subfigure}
\caption{Box plot for the predicted distribution and the real observed counts. We plot the log of relative abundance with zero count samples adjusted to 0.5 in the alcohol and non-alcohol groups for (a) Burkholderiales bacterium and (b) Alistipes indistinctus in Dietary.
}
\label{fig:Example_Dietary_taxa13}
\end{figure}

In Dietary, we present the box plots for Burkholderiales bacterium and Alistipes indistinctus to show the differential variability between the alcohol and non-alcohol drinkers (Figure~\ref{fig:Example_Dietary_taxa13}). For Burkholderiales bacterium, the differential abundance in the two groups (i.e., alcohol and non-alcohol drinkers) are similar, but the overdispersion in alcohol drinkers is obviously larger than that in teetotal subjects. ZIPG can distinguish the differential variability in two groups with $p = 0.00293$ ($H_0 : \beta_1^* = 0$). ZIPG also provides a shorter interquartile in the  non-alcohol drinker group, which is consistent with the raw data. On the contrary, pscl failed to detect any differential abundance ($p = 0.636$), while DESeq2 provided $p = 0.049$ which is significant but much larger than the p-value of ZIPG. For Alistipes indistinctus, though pscl and DESeq2 identified the differential abundance between two groups, both of them did not approximate the data from a distributional perspective because of the ignorance of differential variability. In contrast, ZIPG can distinguish it with $p$ = 8.54e-6 ($H_0 : \beta_1^* = 0$), and provide a long box for the non-alcohol group showing their small overdispersion and a median more closer to real data.

\section{Discussion}\label{s:discussion}
In this paper, we propose a Zero-Inflated Poisson-Gamma model for microbiome count data analysis. We decompose the zero-inflated Poisson model and factor the Poisson mean as $\lambda_{ijk}$ for the average abundance level and a multiplicative factor $U_{ijk}$ following gamma distribution controlled by variation parameter $\theta_{ik}$, which accounts for individual-level microbiome abundance variation around $\lambda_{ijk}$. In traditional ZINB regression, the dispersion parameter is often treated as a nuisance parameter. Our model allows different sets of covariates to be linked to $\lambda_{ijk}$ and $\theta_{ik}$ and provides a valid test, outperforming other negative-binomial-based models such as pscl and NBZIMM. To our knowledge, the ZINB-based Omnibus \citep{chen2018omnibus} method may be the first and only paper that links dispersion to covariates. However, the Omnibus test can not distinguish differential abundance and differential variability. In comparison, we test differential abundance and differential variability separately for longitudinal data and provide valid confidence intervals for each parameter. {\color{black} Moreover, other potential distributions for modeling the multiplicative factor $U_{ijk}$ are worth future exploring, including the mixing distribution of log-normals. However, the difficulty in distinguishing two sources of zeros always exists when the overdispersion is large.}

Though linking $\gamma_k$ to covariates is proposed in pscl and NBZIMM, it is not suggested in our ZIPG model based on two reasons. First, the mechanism of zero-inflation $\gamma_k$ does not have explicit biological interpretation, while $\theta_{ik}$ can be explained as individual-level microbiome stability in longitudinal data. Through simulations, we have shown that linking $\gamma_k$ to covariates is not preferred from the model selection perspective, even if both $\theta_{ik}$ and $\gamma_k$ are covariates-dependent. 
Second, the increment in either $\gamma_k$ or $\theta_{ik}$ will lead to the increment of zeros in observed data, and thus linking both parameters to covariates simultaneously will make the inference more challenging and unreliable. 

Some promising future work could be incorporating auxiliary information from other taxa. One possible way is to assume the taxon-specific dispersion parameters $\theta_{ik}$'s of closely related taxa (e.g., taxa in the same phylogenetic branch) are impacted by covariates $\boldsymbol{X_{i}^*}$ identically and share the same coefficient $\boldsymbol{\beta^*}$. In addition, inference on a group of taxa in a joint multivariate model is also worth future investigation.

\bigskip
\begin{center}
{\large\bf Supplementary Materials}
\end{center}

\begin{description}

\item[Supplements:]
The supplemental materials (ZIPG-appendix.pdf) include  mathematical details of the non-concavity of the log-likelihood, analytical expressions of gradient, details of the parametric bootstrap algorithm, and supplementary figures and tables for additional simulation and real-world data analysis.  

\item[R code for  ZIPG:] The \texttt{R} code for \texttt{ZIPG} is available based on request. We will submit the \texttt{R} package to \texttt{CRAN} once the paper is accepted.


\end{description}

\bibliographystyle{chicago}
\bibliography{JASA.bib}
\end{document}